\documentclass[graybox]{svmult}

\usepackage{mathptmx}       
\usepackage{helvet}         
\usepackage{courier}        
\usepackage{type1cm}        
\usepackage{url,amsmath,amssymb}
\usepackage{makeidx}         
\usepackage{graphicx}        
\usepackage{multicol}        
\usepackage[bottom]{footmisc}

\makeindex             

\begin{document}

\title*{Phases of dense matter in compact stars}
\author{D. Blaschke and N. Chamel}
\institute{D. Blaschke \at Institute of Theoretical Physics, University of Wroclaw, Max Born pl. 9, 50-204 Wroclaw, Poland\\
Bogoliubov Laboratory of Theoretical Physics, joint Institute for Nuclear Research, Joliot-Curie Street 6, 141980 Dubna, Russia\\
National Research Nuclear University (MEPhI), Kashirskoe Shosse 31, 115409 Moscow, Russia, \email{david.blaschke@ift.uni.wroc.pl}
\and N. Chamel \at Institute of Astronomy and Astrophysics, Universit\'e Libre de Bruxelles, CP 226, 
Boulevard du Triomphe, B-1050 Brussels, Belgium, \email{nchamel@ulb.ac.be}}
%
%
\maketitle


\abstract{Formed in the aftermath of gravitational core-collapse supernova explosions, neutron stars are unique cosmic laboratories 
for probing the properties of matter under extreme conditions that cannot be reproduced in terrestrial laboratories. The interior of a neutron star, endowed with the highest magnetic fields known and with densities spanning about ten orders of magnitude from the surface to the centre, is predicted to exhibit various phases of dense strongly interacting matter, whose physics is reviewed in this chapter. The outer layers of a neutron star consist of a solid nuclear crust, permeated by a neutron ocean in its densest region, possibly on top of a nuclear ``pasta'' mantle. The properties of these layers and of the  homogeneous isospin asymmetric nuclear matter beneath constituting the outer core may still be constrained by terrestrial experiments. The inner core of highly degenerate, strongly interacting matter poses a few puzzles and questions which are reviewed here together with perspectives for their resolution. Consequences of the dense-matter phases for observables such the neutron-star mass-radius relationship and the prospects to uncover their structure with modern observational programmes are touched upon.
}

\section{Introduction}
\label{sec:intro}

\subsection{Cosmic laboratories}
\label{sec:lab}

Neutron stars are the stellar remnants of massive stars at the end point of their evolution (see, e.g., Ref.~\cite{hae07}). Neutron stars have a 
 mass between one and two times that of the Sun, but packed into a space only 20 km across (100,000 times smaller than the Sun's diameter). The average density of a neutron star can thus exceed a few hundred-thousand billion grams per cubic centimeter - a density higher than that found inside the heaviest  atomic nuclei. Neutron stars are not only the most compact observed stars in the Universe, but they are also endowed with the strongest magnetic fields known, which could reach millions of billions times that of the Earth. Neutron-star observations thus offer the unique opportunity to explore the properties of matter under extreme conditions, which cannot be reproduced in the laboratory.

\subsection{Cold catalysed matter hypothesis}
\label{sec:catalysed}

During the formation of a neutron star, the hot compressed matter in the collapsed stellar core is assumed to undergo all kinds of nuclear and electroweak processes 
such that the compact stellar remnant cools down by following a sequence of full thermodynamic quasi equilibrium states. The resulting neutron star thus eventually consists of 
``cold catalysed matter''\index{catalysed matter}, i.e., electrically charge neutral matter in its absolute ground state, after all internal heat has been released~\cite{hw58,htww65}. This scenario supposes that the reaction rates are 
much higher than the cooling rate. In reality, the composition of a neutron star may not only depend on the particular internal conditions prevailing during its formation, 
but also on its subsequent evolution, as well as on its environment. In particular, the constitution of the collapsed core may become ``frozen in'' as it cools down (see, e.g. Ref.~\cite{goriely2012} for a recent discussion). 
Moreover, the accretion of matter from a companion star or a high enough magnetic field may notably change the composition of the neutron star. These different situations will be separately discussed.

\section{Surface layers of a neutron star}
\label{sec:surface}

Neutron stars are expected to be surrounded by a very thin atmosphere consisting of a plasma of electrons and light elements (mainly hydrogen and helium 
though heavier elements like carbon may also be present~\cite{ho09}). Its properties such as the effective temperature, the composition, and the magnetic field
configuration, can be inferred by analyzing the thermal X-ray emission from neutron stars (see, e.g. Ref.~\cite{potekhin2015}). The region beneath consists 
of a solid crust (see, e.g., Ref.~\cite{lrr}).

\subsection{Nonaccreted neutron stars}
\label{sec:surface-nonaccreted}

The outermost region of a nonaccreting neutron star is expected to be made of iron $^{56}$Fe, the end-product of stellar nucleosynthesis. 
The properties of compressed iron can be probed in terrestrial laboratories up to pressures of order $10^{14}$~dyn~cm$^{-2}$ with nuclear explosions and 
laser-driven shock-wave experiments (see, e.g., Refs.~\cite{bata2002,fort2010,ping2013}). Under these conditions, iron has an hexagonal close-packed 
structure~\cite{ping2013,stixrude2012}. Transitions to a face-centred cubic lattice and a body-centred cubic lattice are expected at pressures 
of about $6\times 10^{13}$~dyn~cm$^{-2}$ and $4\times 10^{14}$~dyn~cm$^{-2}$ respectively for temperatures below $\sim 10^4-10^5$~K~\cite{stixrude2012}. 
Although such pressures are tremendous according to terrestrial standards, they still remain negligibly small compared to those prevailing in a neutron star (note also that observed middle-aged neutron stars have typical surface temperatures of order $10^6$~K, see e.g. Ref.~\cite{potekhin2018}). 
In particular, the highest density of iron that has been experimentally attained is only about three times the density at the surface of a cold neutron star, 
and 14 orders of magnitude lower than the density at the stellar center. This corresponds to a depth of about $0.1$~mm for a star with a mass 
$\mathcal{M}=1.4 \mathcal{M}_\odot$ and a radius $R=12$~km\footnote{The depth $z_0$ below the surface of a cold and fully catalysed neutron star at a pressure $P_0$ was estimated as 
$\displaystyle z_0\approx \int_{0}^{P_0} \frac{dP}{\rho g_s}$, with $\displaystyle g_s=\frac{G\mathcal{M}}{R^2} \left(1-\frac{2 G\mathcal{M}}{R c^2}\right)^{-1/2}$, 
see, e.g., Ref.~\cite{lrr}, and we have made use of an interpolation of the ``QEOS'' equation of state from  Ref.~\cite{more88}, as tabulated in Ref.~\cite{lai91}.}. Deeper in the 
star, recourse must therefore be made to theoretical models (see, e.g. Ref.~\cite{lai91} and references therein). At a density $\rho_{\rm eip}\approx 2\times 
10^4$~g~cm$^{-3}$ (about $22$~cm below the surface\footnote{The depth was calculated combining the equations of state labeled ``QEOS'' from Ref.~\cite{more88} and ``TFD'' in Table 5 
of Ref.~\cite{lai91}.}), the interatomic spacing $a_N=[3/(4\pi n_N)]^{1/3}$ (with $n_N$ the number density of atomic nuclei) becomes comparable with the atomic 
radius $\sim a_0/Z^{1/3}$ (with $a_0$ the Bohr radius and $Z$ the atomic number - here $Z=26$). At densities $\rho \gg \rho_{\rm eip}$, atoms are crushed into 
a dense plasma of nuclei and free electrons\index{Coulomb plasma}. This ionization is not triggered by thermal excitations but by the prodigious gravitational pressure (see, e.g., 
Ref.~\cite{hae07}). Beyond this point, electrons are very weakly perturbed by ions, and thus behave as an essentially ideal Fermi gas. Because electrons are 
highly degenerate, they provide the necessary pressure to counterbalance the weight of the layers above. 
The attractive Coulomb interactions between electrons and ions reduce the electron Fermi gas pressure, but their relative contribution 
is only of order $Z^{2/3} e^2/(\hbar v_F)$~\cite{hae07} (with $e$ the elementary electric charge, $\hbar$ the Planck-Dirac constant, and $v_F$ the electron 
Fermi velocity). The (electric charge) polarization of electrons around ions (also referred to as electron screening) leads to a correction of order 
$Z^{4/3} e^4/(\hbar v_F)^2$~\cite{hae07}. For the nuclides present in the crust, the electron exchange correction of order $e^2/(\hbar v_F)$ turns out be 
of the same magnitude. For a discussion of higher-order corrections due to electron correlations, finite size of nuclei, and quantum-zero point motion of nuclei, 
see e.g. Refs.~\cite{guo2007,pearson2011}. Although electron-ion interactions are very small, they still play a major role in the equilibrium structure of the crust. 
As the density reaches about $7\times10^6$~g~cm$^{-3}$ (about $11$~m below the surface\footnote{The depth was calculated combining the 
equations of state labeled 
``QEOS'' and ``TFD'' in Table 5 of Ref.~\cite{lai91}, and the equation of state calculated in Ref.~\cite{pearson2011}.}), electrons become relativistic 
since the interelectron spacing $a_e=[3/(4\pi n_e)]^{1/3}$ (with $n_e$ the electron number density) is comparable with the electron Compton 
wavelength $\lambda_e=\hbar/(m_e c)$ ($m_e$ is the electron mass, and $c$ is the speed of light). 

\subsection{Accreted neutron stars}
\label{sec:surface-accreted}

The composition of the surface layers of a neutron star may be changed by the fallback of material from the envelope ejected during the supernova explosion, and more 
importantly by the accretion\index{accretion} of matter from a stellar companion (see, e.g., Ref.~\cite{lrr}). 
The accreted material forms an hydrogen rich envelope around the star. Stable hydrogen burning produces helium, which accumulates in a layer beneath. These helium 
ashes 
ignite under specific conditions of density (typically $\sim 10^6-10^7$ g~cm$^{-3}$) and temperature. For some range of accretion rates, 
helium burning is unstable, converting within seconds all the envelope into nuclides in the nickel-cadmium range~\cite{schatz2001}. These thermonuclear explosions are 
observed as X-ray bursts\index{X-ray bursts}, with luminosity 
up to about $10^{38}~{\rm erg~s^{-1}}$ ($\approx$ Eddington limit for neutron stars), and with a typical 
decay lasting a few tens of seconds. Multiplying the burst luminosity by its duration we get an estimate of the total burst energy $\sim 10^{39}-10^{40}$ erg. 
X-ray bursts are quasiperiodic, with typical recurrence time of about hours to days. Less frequent but more energetic ($\sim 10^{42}$ erg) are superbursts lasting 
for a few hours, with 
recurrence times of several years. These superbusts are presumably triggered 
by the unstable burning of carbon at densities $\sim 10^8-10^9$ 
g~cm$^{-3}$, and their ashes are predicted to consist mainly of $^{66}$Ni, $^{64}$Ni, $^{60}$Fe, and $^{54}$Cr~\cite{schatz2003}. 

Because the equilibrium structure of such multicomponent plasmas\index{Coulomb plasma} is highly uncertain, various properties of accreted neutron-star crusts such as their breaking strain 
remain poorly known. According to classical molecular dynamics simulations, ashes tend to arrange on a regular body-centred cubic lattice as in catalysed crusts~\cite{horowitz2009}. 
However, the system may have not fully relaxed to its true equilibrium state due to the very slow dynamics of crystal growth. Alternatively, valuable insight can 
be gained from analyses of the dynamical stability of given crystal structures (but with arbitrary composition)~\cite{kozhberov2015}. The reduced 
computational cost allows to take into account quantum effects, which can play a key role at low temperatures. Recently, a different approach has been followed using 
genetic algorithms~\cite{engstrom2016}. Although limited to a few ternary crystals, unconstrained global searches of the equilibrium structure have revealed a very 
rich phase diagram with noncubic lattices. In view of the importance of the crust structure, these theoretical studies should be pursued. Astrophysical observations 
can provide complementary information. In particular, the possibility of an amorphous crust has been ruled out by cooling simulations of the observed thermal relaxation 
of transiently accreting neutron stars~\cite{shternin2007}. Gravitational-wave observations (or lack thereof) may provide additional information on the structure of neutron-star crusts~\cite{hoffman2012}. 

\subsection{Highly magnetised neutron stars} 
\label{sec:surface-magnetars}

Neutron stars are not only the most compact observed stars in the universe, but are also among the strongest magnets known, with  
typical surface magnetic fields of order $10^{12}$~G~\cite{seir04}. A few radio pulsars have been recently found to have significantly 
higher surface magnetic fields of order $10^{13}-10^{14}$~G~\cite{ka11}. Surface magnetic fields of order $10^{14}-10^{15}$~G have 
been inferred in soft-gamma ray repeaters (SGRs) and anomalous x-ray pulsars (AXPs) from both spin-down and spectroscopic studies~\cite{mcgill14,tien13,hong14}. Various observations 
suggest that the interior magnetic field may be even higher~\cite{ste05,kam07,vie07,rea10,maki14}. At the time of this 
writing, 11 SGRs and 12 AXPs have been already identified~\cite{mcgill14}. It is now widely believed that these objects belong 
to a different class of neutron stars called \emph{magnetars}\index{magnetar} (see e.g. Ref.~\cite{wt06} for a review), as proposed by Duncan 
and Thomson in 1992~\cite{td92}. Numerical simulations confirmed that magnetic fields of order $\sim 10^{15}-10^{16}$~G can be 
produced during supernovae explosions due to the magnetorotational instability~\cite{ard05}. Theoretical considerations corroborated by numerical simulations suggest that neutron stars may potentially 
possess internal magnetic fields as high as $10^{18}$~G (see, e.g. Refs.~\cite{kiuchi2008,frieben2012,pili14,chatterjee2015} and references therein). 

The properties of the surface layers of a neutron star can be drastically different in the presence of a high magnetic field. 
The electron motion perpendicular to the magnetic field lines is quantised into Landau orbitals\index{Landau quantisation} with a characteristic magnetic length scale 
(see, e.g., Ref.~\cite{hae07}) $a_m=a_0\sqrt{B_{\rm at}/B}$, with
\begin{equation}
B_{\rm at}=\frac{m_e^2 e^3 c}{\hbar^3} \simeq 2.35\times 10^9~{\rm G}\, . 
\end{equation}
For magnetic fields $B\gg B_{\rm at}$, atoms are expected to adopt a very elongated shape along the magnetic field lines and to form linear chains. 
The attractive interaction between these chains could lead to a phase transition into a magnetically condensed phase with a surface density 
estimated as~\cite{lai91b}
\begin{equation}
 \rho_s \simeq 560 A Z^{-3/5} B_{12}^{6/5}\, {\rm g~cm}^{-3}\, ,
\end{equation}
where $B_{12}\equiv B/(10^{12})$~G (assuming $B\ll 10^{18}$~G, see e.g. Ref.~\cite{chapav12}). 
For iron with $B=10^{15}$~G, we obtain $\rho_s\simeq 1.8\times 10^7$ g~cm$^{-3}$, as compared 
to $7.86$ g~cm$^{-3}$ in the absence of magnetic fields. In deeper regions of the crust, the density $\rho$ at pressure $P$ is approximately 
given by~\cite{chapav12}
\begin{equation}
\rho \approx \rho_s \left(1+\sqrt{\frac{P}{P_0}}\right)\, ,
\end{equation}
where 
\begin{equation}
P_0 \simeq 1.45\times 10^{20} B_{12}^{7/5}\left(\frac{Z}{A}\right)^2 \, {\rm dyn\, cm}^{-2}\, .
\end{equation}
The presence of a high magnetic field does not change the equilibrium structure of a body-centred cubic crystal of ions embedded in a 
uniform charge neutralizing electron background~\cite{kozhberov2016}. However, it has been found that in the strongly quantising regime, 
the long-range part of the ion-ion potential exhibits Friedel oscillations that may lead to the formation of strongly coupled filaments 
aligned with the magnetic field~\cite{bedaque2013}. This possibility should be further examined. 

The absence of spectral features in the thermal emission from seven radio-quiet isolated neutron stars (usually referred to as the 
'Magnificent Seven') could be the consequence of the magnetic condensation of their surface~\cite{turolla2004}, with magnetic fields of order 
$10^{13}-10^{14}$~G as estimated from X-ray timing data. The presence of high magnetic fields has recently found additional support 
from optical polarimetry measurements~\cite{mignani2017}. Future X-ray polarimetry measurements could potentially allow to discriminate 
between the case of a gaseous atmosphere and a condensed magnetic surface.

\section{General considerations on dense stellar plasmas}
\label{sec:plasmas}

In this section, we consider fully ionised Coulomb plasmas\index{Coulomb plasma} at densities $\rho$ above $10^7$ g~cm$^{-3}$ and below the onset of neutron emission by nuclei. 

\subsection{Gravitational stratification} 
\label{sec:plasmas-gravitation}

Because the pressure has to vary continuously 
throughout the star and the nucleon number is conserved, the suitable thermodynamic potential for determining the composition is the Gibbs free energy per 
nucleon $g$~\cite{tondeur71,bps71}. As shown in the appendix of Ref.\cite{chafant15}, $g$ remains the suitable thermodynamic potential in the presence 
of a high magnetic field. 
At a given pressure, hot dense matter in full thermodynamic equilibrium generally consists of an admixture of 
various nuclear species~\cite{hempel2012,furusawa2013,buy2014,gulm2015,grams2018}. As the temperature decreases, the distribution of nuclei becomes very narrow. 
The crust of a cold nonaccreted neutron star is thus expected to be stratified into different layers, each of which consists of a body-centred cubic crystal 
made of a single nuclear species. 
Substitutional binary ionic compounds with cesium chloride structure can only possibly exist at the interface between two adjacent strata~\cite{cf2016}. 
Multinary ionic compounds may however be present in the crust of accreted neutron stars~\cite{cha2017}. We shall not discuss this possibility here. 
In the single-nucleus approximation, each layer is described as a one-component crystal of pointlike nuclei in a uniform charge neutralizing electron background. 
Retaining only the electrostatic correction to the ideal electron Fermi gas model, and expanding $g$ to first order in $\alpha=e^2/(\hbar c)$, the transition 
between two adjacent strata made of nuclei $(A_1,Z_1)$ and $(A_2,Z_2)$ respectively, is determined by the condition~\cite{cf2016} 
\begin{equation}\label{eq:threshold-condition}
\mu_e + C\, \alpha \hbar c n_e^{1/3}\biggl(\frac{4}{3}\frac{Z_1^{5/3}}{A_1} - \frac{1}{3}\frac{Z_1^{2/3} Z_2}{A_2} -\frac{Z_2^{5/3}}{A_2}\biggr)
\left(\frac{Z_1}{A_1}-\frac{Z_2}{A_2}\right)^{-1} = \mu_e^{1\rightarrow2}\, ,
\end{equation}
where $\mu_e$ is the electron Fermi energy, $M^\prime(A_1,Z_1)$ and $M^\prime(A_2,Z_2)$ are the nuclear masses, $C$ is a dimensionless structure constant (see discussion below), and 
\begin{equation}\label{eq:thres-mue}
 \mu_e^{1\rightarrow2}\equiv \biggl[\frac{M^\prime(A_2,Z_2)c^2}{A_2}-\frac{M^\prime(A_1,Z_1)c^2}{A_1}\biggr]\left(\frac{Z_1}{A_1}-\frac{Z_2}{A_2}\right)^{-1} +  m_e c^2\, .
\end{equation}
 
According to the Bohr-van Leeuwen theorem~\cite{bvl32}, the electrostatic correction is independent of the magnetic field apart from a negligibly small 
contribution due to quantum fluctuations of ion motion~\cite{bai09}. Equations~(\ref{eq:threshold-condition}) and (\ref{eq:thres-mue}) thus remain valid 
for highly magnetised neutron stars (provided the electrostatic correction remains small\footnote{Because the electron chemical potential scales as $\mu_e\sim m_e c^2 \lambda_e^3 n_e/B$ in strongly quantising magnetic fields, the expansion of $g$ to first order in $\alpha$ eventually breaks down in high enough magnetic fields. See also Section~\ref{sec:outer-crust-magnetar}.}), but nuclear masses hence $\mu_e^{1\rightarrow2}$ could change substantially (especially if the magnetic field strength exceeds $10^{17}$~G) compared to their values 
in the absence of a magnetic field~\cite{pen11,stein16}. 
As shown in section~\ref{sec:plasmas-neutron-drip}, mechanical stability requires $Z_1/A_1 > Z_2/A_2$ so that Eqs.~(\ref{eq:threshold-condition}) and (\ref{eq:thres-mue}) 
are well defined. 
We adopt here the convention to include the rest mass of $Z$ electrons in $M^\prime(A,Z)$. The reason is that experimental \emph{atomic} masses  $M(A,Z)$ are generally 
tabulated rather than \emph{nuclear} masses $M^\prime(A,Z)$. The latter can be obtained from the former after subtracting out the binding energy of the atomic electrons 
(see, e.g., Eq.~(A4) of Ref.~\cite{lpt03}). The threshold condition~(\ref{eq:threshold-condition}) still holds in hot dense matter whether the Coulomb plasma is in a 
solid or a liquid state, provided the temperature $T$ is well below the electron Fermi temperature defined by 
\begin{equation}
 T_{{\rm F}e}=\frac{\mu_e-m_e c^2}{k_{\rm B}}\approx 5.93\times 10^9 \frac{\mu_e}{m_e c^2}~{\rm K}\, .
\end{equation}
In principle, nuclei in hot dense matter may coexist with a gas of nucleons and light particles, but their fraction is negligibly small at $T\ll T_{{\rm F}e}$ 
(see, e.g., Ref.~\cite{hae07}). The structure constant is very-well approximated by the ion-sphere model~\cite{salpeter54}
\begin{equation}\label{eq:WS-approx}
C=-\frac{9}{10}\left(\frac{4\pi}{3}\right)^{1/3}\, ,
\end{equation}
which also provides a lower bound~\cite{lieb75}. 
The structure constant of the solid phase (assuming a perfect body-centred cubic crystal) is $C_{\rm bcc}\simeq -1.4442$, whereas in the liquid phase $C_{\rm liq}\simeq -1.4621$ (see, e.g., Ref.~\cite{hae07}).
Crystallization of a one-component plasma of ions with charge $Z$ occurs at the temperature given by (see, e.g., Ref.~\cite{hae07})
\begin{equation}\label{eq:Tm}
T_m=\frac{e^2}{a_e k_{\rm B} \Gamma_m} Z^{5/3} \, ,
\end{equation}
where $k_{\rm B}$ is Boltzmann's constant, and $\Gamma_m$ is the Coulomb coupling parameter at melting. In the absence of magnetic fields, $\Gamma_m\simeq 175$~\cite{hae07}. In the presence of a high magnetic field, Coulomb crystals are expected to be more 
stable so that $\Gamma_m\lesssim 175$~\cite{pot13}. 
Transitions between multicomponent phases as in accreted neutron-star crusts have been addressed in Ref.~\cite{cf2016}.

\begin{figure}
\sidecaption[t]
 \includegraphics[scale=0.7]{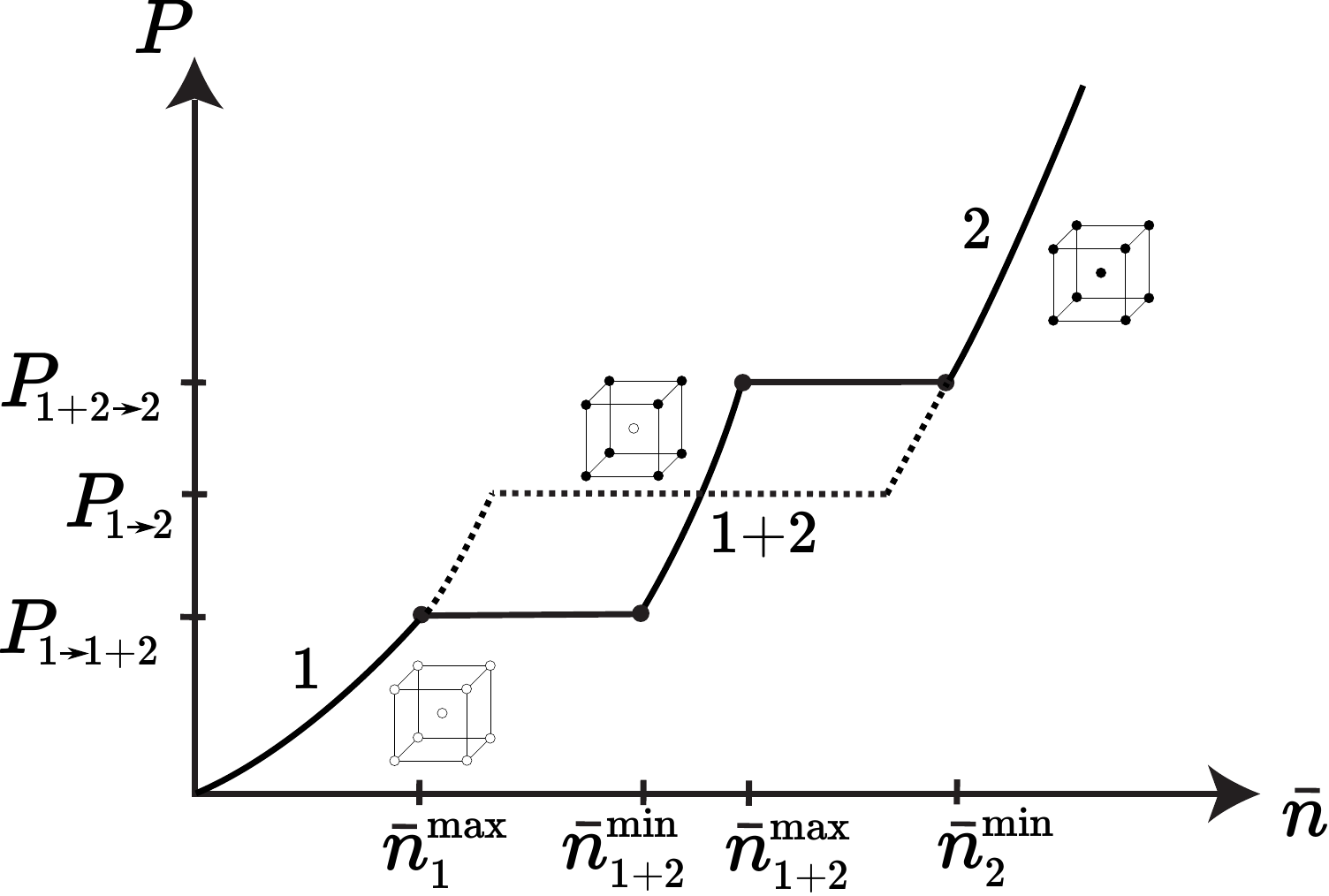}
 \caption{Schematic representation of the pressure $P$ versus mean baryon number density $\bar n$ for a transition between two pure body-centred cubic solid phases 
 of nuclei $(A_1,Z_1)$ and $(A_2,Z_2)$ accompanied by the formation of a substitutional binary compound. For comparison, the transition leading to the coexistence 
 of pure phases is indicated by the dotted line. The figure is not to scale. Taken from Ref.~\cite{cf2016}. 
  \label{fig:transit}
 }
\end{figure}

Because the transition between two adjacent layers occurs at a fixed pressure, it is accompanied by a density discontinuity  given by~\cite{cf2016} 
\begin{eqnarray}\label{eq:density-jump}
\frac{\bar n_{2}^{\rm min}- \bar n_1^{\rm max}}{\bar n_1^{\rm max}} =  \frac{A_2}{Z_2}\frac{Z_1}{A_2} \Biggl[1+
\frac{1}{3}C \alpha\hbar c \left(n_e^{2/3} \frac{d\mu_e}{dn_e} \right)^{-1} \biggl(Z_1^{2/3}-Z_2^{2/3}\biggr)\Biggr] -1\, ,
\end{eqnarray}
as schematically illustrated in Fig.~\ref{fig:transit}. It is to be understood that the electron density is obtained from Eq.~(\ref{eq:threshold-condition}). 
In principle, pure phases of nuclei ($A_1,Z_1$) and ($A_2,Z_2$) can coexist at intermediate densities. However, since inside a self-gravitating body in 
hydrostatic equilibrium the pressure must increase monotonically with depth (see, e.g., Ref.~\cite{hae07}), such coexisting phases cannot be present in the crust 
of a neutron star. Instead, binary compounds could form at the boundary but only over a small range of pressures 
$(P_{1+2\rightarrow2} - P_{1\rightarrow1+2})/ P_{1\rightarrow2} \ll 1$, and provided $Z_1\neq Z_2$~\cite{cf2016}. As a matter of fact, transitions from a pure 
phase to a compound phase is still accompanied by density jumps. Very accurate analytical expressions for the pressure at the transition 
between two adjacent strata as well as the average baryon number densities of each layer in the absence of magnetic field can be found in Ref.~\cite{cf2016}. 
The errors were found to lie below 0.2\% for the pressures, and below 0.07\% for the densities. Approximate expressions in the presence of a strongly 
quantising magnetic field are given in Ref.~\cite{chamel2017b}.

\subsection{Matter neutronization and onset of neutron emission}
\label{sec:plasmas-neutron-drip}

As a consequence of Le Chatelier's principle, the bulk modulus $K=\bar n dP/d\bar n$ of matter in equilibrium must be positive: the density must therefore 
increase with pressure. 
Using Eq.~(\ref{eq:density-jump}) and ignoring the small electrostatic correction, we conclude that nuclei become progressively more neutron rich with increasing depth ($Z_2/A_2 < Z_1/A_1$) (see also Ref.~\cite{blasio2000} where the restoring force acting on displaced ions was explicitly calculated). This neutronization is achieved by electron captures, and various nuclear processes (e.g. fissions, fusions) depending on the local conditions. In turn, 
assuming that the electrostatic contribution in the left-hand side of Eq.~(\ref{eq:threshold-condition}) is small\footnote{This assumption may be violated in the 
presence of a very high magnetic field, as discussed in Section~\ref{sec:plasmas-gravitation}.}, $\mu_e^{1\rightarrow2}$ must be positive implying that $M^\prime(A_2,Z_2)/A_2 > M^\prime(A_1,Z_1)/A_1$. Combining these
two inequalities, we find that as the composition changes nucleons become less bound $B(A_2,Z_2)/A_2 < B(A_1,Z_1)/A_1$, where $B(A,Z)=Z m_p c^2 + (A-Z) m_n c^2 
- M^\prime(A,Z) c^2$ denotes the total binding energy of the nucleus ($A,Z$). At some point, $\Delta N>0$ neutrons will become unbound and start to drip out 
of nuclei. This transition marks the boundary between the outer and inner regions of the crust. 
Due to baryon number conservation, the daughter nuclei will be of the form ($A-\Delta N,Z-\Delta Z$), with $Z\geq \Delta Z\geq \Delta N + Z -A$ 
(this follows from the requirement that the daughter nuclei must contain positive numbers of neutrons and protons). 
Assuming $\Delta Z\neq 0$, and ignoring neutron band-structure effects, the onset of this transition is determined by the condition~\cite{cfzh15}
\begin{equation}\label{eq:neutron-drip}
\mu_e + C \alpha \hbar c n_e^{1/3}\biggl[\frac{Z^{5/3}-(Z-\Delta Z)^{5/3}}{\Delta Z} + \frac{1}{3} Z^{2/3}\biggr] =  \mu_e^{\rm drip} \, ,
\end{equation}
where
\begin{equation}\label{eq:muebetan}
\mu_e^{\rm drip}\equiv \frac{M^\prime(A-\Delta N,Z-\Delta Z)c^2-M^\prime(A,Z)c^2 +m_n c^2 \Delta N}{\Delta Z} + m_e c^2 \, .
\end{equation}

In the case $\Delta Z=0$ and $\Delta N>0$, the threshold condition reads 
\begin{equation}
M^\prime(A,Z)- M^\prime(A-\Delta N, Z)  = \Delta N m_n \, ,
\end{equation}
independently of the electron background: this shows that a nucleus unstable against neutron emission in vacuum is also unstable in a 
stellar environment at \emph{any} density. This means in particular that the neutron-drip transition in the crust of a neutron star must be 
necessarily triggered by reactions such that $\Delta Z \neq 0$. Since $\mu_e^{\rm drip}>m_e c^2$, the nucleus must satisfy the following 
constraint 
\begin{equation}\label{eq:drip-constraint}
M^\prime(A-\Delta N,Z-\Delta Z)-M^\prime(A,Z) +m_n \Delta N > 0 \, .
\end{equation}

Equations~(\ref{eq:threshold-condition}), (\ref{eq:density-jump}), and (\ref{eq:neutron-drip}) are applicable to both weakly and highly magnetised 
neutron stars (with suitable values for the nuclear masses), accreted and nonaccreted (to the extent that the crust consists of pure layers). 
Very accurate analytical expressions for the pressure and the density at the neutron-drip transition\index{neutron drip} can be found in Refs.~\cite{cfzh15,chasto15}.

\section{Outer crust of a neutron star}
\label{sec:outer-crust}
\index{outer crust}
\subsection{Nonaccreted neutron stars}
\label{sec:outer-crust-nonaccreted}

The internal constitution of the outermost layers of the crust of a nonaccreted (fully catalysed\index{catalysed matter}) neutron star is completely determined by experimental atomic mass measurements. In particular, 
the strata of iron $^{56}$Fe is predicted to extend up to a density of about $8\times 10^6$~g~cm$^{-3}$ (about $12$~m below the surface\footnote{The depth was calculated combining the equations of state labeled ``QEOS'' and ``TFD'' in Table 5 of Ref.~\cite{lai91}, and the equation of state calculated in Ref.~\cite{pearson2011}.}). 
Using the data from the 
2016 Atomic Mass Evaluation (AME)~\cite{audi16} and applying the model of Ref.~\cite{pearson2011}, the region beneath is found to consist of a succession of layers 
made of $^{62}$Ni, $^{64}$Ni, $^{66}$Ni, $^{86}$Kr, $^{84}$Se, $^{82}$Ge, and $^{80}$Zn with increasing depth. Although electron charge polarisation effects are very 
small, they may change the composition (see, e.g., Ref.~\cite{cf2016a}). For instance, ignoring this correction leads to the appearance of $^{58}$Fe (whose 
mass is experimentally known) in between the layers of $^{62}$Ni and $^{64}$Ni. This nuclide is still present if the electron polarisation is implemented
using the interpolating formula of Ref.~\cite{pot00} instead of the limiting Thomas-Fermi expression for $Z\rightarrow +\infty$~\cite{salpeter61} employed 
in Ref.~\cite{pearson2011}. The existence or not of this nuclide in the crust is also found to depend on the precision of the nuclear mass measurements: using the 
data from the 2012 AME~\cite{audi12}, which differs by less than 1 keV/$c^2$ compared to the more recent value~\cite{audi16}, $^{58}$Fe thus disappears independently 
on how the electron polarisation correction is calculated. 

The composition of the innermost regions is more uncertain due to the lack of experimental data, and 
can only be explored using nuclear mass models (see, e.g., Refs.~\cite{guo2007,pearson2011,rust2006,roca2008,chamel2015c,bcpm,utama2016,fantina2017,chamel2017} 
for recent calculations). The most accurate nuclear mass models achieve to fit all known experimental masses with a root-mean square deviation of a few hundred 
keV/$c^2$~\cite{pearson2013,sobiczewski2014}, whereas errors of the order of a keV/$c^2$ can change the composition as previously discussed. Therefore, the main sources of uncertainties 
are nuclear masses. Because of nuclear pairing effects, nuclei with even numbers of neutrons and protons are more tightly bound that neighboring nuclei in the nuclear
chart. For this reason, odd nuclei were generally not considered in earlier calculations of neutron-star crusts, and the search for the equilibrium nuclides was limited 
to a restricted set of even-even nuclei (130 even-even nuclei in the seminal work of Ref.~\cite{bps71}). However, the crustal composition is determined by the full 
thermodynamic equilibrium with respect to all processes, not only strong nuclear reactions. Therefore, there is no fundamental reason to rule out odd nuclei. As a 
matter of fact, the microscopic nuclear mass model HFB-21 predicts the presence of $^{79}$Cu and $^{121}$Y in the outer crust~\cite{pearson2011}. 
The nuclear shell structure has a profound influence on the crustal composition, with a predominance of nuclides with neutron magic numbers $N=50$ and $N=82$. 
Because of the requirement of $\beta$ equilibrium and electric charge neutrality, the proton number $Z$ is more tightly constrained and for this reason, only a few crustal layers are made of 
nuclei with proton magic number $Z=28$, and the only doubly magic nucleus predicted by various nuclear mass models to be present in the crust is $^{78}$Ni~\cite{xu14,hagen2016b}. 

Progress in experimental techniques over the past decades have allowed to probe deeper the interior of a neutron star crust, up to a density $\sim 6 \times 10^{10}$~g~cm$^{-3}$. 
For comparison, the most exotic experimentally measured nuclide predicted to exist in the crust of a neutron star in 1971 was $^{84}$Se ($Z/A\simeq 0.405$) at 
densities up to about $8.2\times 10^9$~g~cm$^{-3}$~\cite{bps71}. 
More recently, the presence of the neutron-rich zinc isotope $^{82}$Zn ($Z/A\simeq 0.366$) that 
was predicted by some nuclear mass models has been ruled out by experiments at the ISOLDE-CERN facility~\cite{wolf13,hemp2013}. In contrast to the modeling of 
gravitational core-collapse supernova explosions that require the knowledge of the properties of a very large ensemble of nuclei, significant advances in the 
understanding of the crust of a nonaccreted neutron star could thus be achieved in the near future by measurements of a few exotic nuclei. Experiments could also provide 
crucial information on the evolution of the nuclear-shell structure towards the neutron-drip line (see, e.g. Ref.~\cite{stepp13} and references therein). 
As discussed in Section~\ref{sec:plasmas-neutron-drip}, the nuclei $(A,Z)$ that could possibly exist in the outer crust of a neutron star beneath the layer of $^{80}$Zn 
must satisfy the \emph{experimental} constraints
\begin{equation}\label{eq:exp-outer-crust}
 Z/A < 0.375, \hskip1cm  M^\prime(A,Z)/A> 930.848~{\rm MeV}/c^2\, ,
\end{equation}
where we have made use of the latest mass data from Ref.~\cite{audi16}. These conditions apply to \emph{all} regions of the outer crust below that 
containing $^{80}$Zn. The conditions~(\ref{eq:exp-outer-crust}) rule out the doubly magic nuclei $^{48}$Ca, $^{48}$Ni, and $^{56}$Ni since $Z/A\simeq 0.417,0.583$ 
and 0.5 respectively. On the other hand, the doubly magic nucleus $^{78}$Ni ($Z/A\simeq 0.359$) is not necessarily excluded. 

Under the assumption of cold catalysed matter, the neutron drip transition is determined considering all possible electron capture and neutron emission processes
with all possible values of $\Delta Z$ and $\Delta N$. The lowest threshold pressure is reached for $\Delta Z=Z$ and $\Delta N=A$ (see e.g. Ref.~\cite{cfzh15}). 
In this case, Eq.~(\ref{eq:drip-constraint}) reduces to $M^\prime(A,Z)/A < m_n$: this condition is satisfied by any nucleus and therefore does not provide any 
additional constraint. Using microscopic nuclear mass models, the neutron-drip density and pressure are predicted to lie in the range 
$\rho_{\rm drip}\sim 4.2-4.5\times 10^{11}$ g~cm$^{-3}$, and $P_{\rm drip}\sim 7.7-8.0\times 10^{29}$ dyn~cm$^{-2}$ respectively~\cite{cfzh15,fantina2016}. 
As discussed in Section~\ref{sec:plasmas-neutron-drip}, the equilibrium nucleus found by minimising the Gibbs free energy per nucleon must be stable 
against neutron emission. Still, it may lie beyond the ``neutron-drip line'' in the chart of nuclides. Let us recall that this line is generally defined at each value 
of the proton number $Z$ by the lightest isotope for which the neutron separation energy $S_n$, defined by
$S_n(A,Z)\equiv M^\prime(A-1,Z)c^2-M^\prime(A,Z)c^2+m_n c^2$
is negative, i.e., the lightest isotope that is unstable with respect to the emission of one neutron. Because of pairing and shell effects,
many nuclei beyond the neutron-rich side of the neutron drip line are actually stable. The outer crust of a neutron star may actually contain ultradrip nuclei. 
For instance, the HFB-21 mass model predicts the presence of $^{124}$Sr with $S_n=0.83$ MeV; the isotope at the neutron-drip line is $^{121}$Sr with 
$S_n=-0.33$ MeV~\cite{pearson2011}.

\subsection{Highly magnetised neutron stars}
\label{sec:outer-crust-magnetar}

According to the magnetar theory\index{magnetar}, neutron stars are born with very high magnetic fields of order $B\sim 10^{16}-10^{17}$~G. The presence of 
such high magnetic fields may alter the formation of the crust. 

As discussed in Section~\ref{sec:surface-magnetars}, the motion of electrons is quantised into Landau orbitals\index{Landau quantisation} in the presence of a high magnetic field. 
If the magnetic field strength exceeds the value 
\begin{equation}
\label{eq:Brel}
B_{\rm rel}=\frac{m_e^2 c^3}{e\hbar}\simeq 4.41\times 10^{13}\, \rm G\, , 
\end{equation}
the orbital size $a_m$ is lower than the electron Compton wavelength $\lambda_e$, so that the electron motion is relativistic.  This situation is 
encountered in SGRs, AXPs, as well as in some radio pulsars. The energy levels of a relativistic electron gas in a magnetic field were first calculated 
by Rabi~\cite{rab28}. 
The quantising effects of the 
magnetic field on the properties of the outer crust of a neutron star are most important when only the first Rabi level is occupied. In such case, 
the magnetic field is usually referred to as strongly quantising. This situation arises when $\rho<\rho_{B}$ and $T<T_B$ with 
\begin{equation}\label{eq:neB}
\rho_{B}=\frac{A}{Z} m \frac{B_\star^{3/2}}{\sqrt{2} \pi^2 \lambda_e^3}\simeq 2.07\times 10^6 \frac{A}{Z} B_\star^{3/2}~{\rm g~cm}^{-3} \, ,
\end{equation}
\begin{equation}\label{eq:TB}
 T_B=\frac{m_e c^2}{k_{\rm B}} B_\star\simeq 5.93\times 10^9 B_\star~\rm K\, ,
\end{equation}
where $B_\star\equiv B/B_{\rm rel}$ and $m=M^\prime(A,Z)/A$ is the mean mass per nucleon.

For the ``low'' magnetic fields $B_\star \lesssim 1$ prevailing in most neutron stars, the internal constitution of their outer crust is essentially the 
same as in the absence of magnetic fields except possibly near the stellar surface, as discussed in Section~\ref{sec:surface-magnetars}. For higher magnetic 
fields $B_\star\gg 1$ as measured in SGRs, AXPs, and some radio pulsars, the composition is found to depend on the magnetic field~\cite{lai91b,
chapav12,nandi2011,chapav13,nandi2013}. In particular, the maximum baryon number density $\bar n^{\rm max}$ up to which a nuclide $(A_1,Z_1)$ is 
present exhibit typical quantum oscillations as a function of the magnetic field strength~\cite{chamel2017b,chasto15,chamut16}. Beyond some magnetic field strength 
$B_\star^{1\rightarrow2}\approx 1/2 (\mu_e^{1\rightarrow2}/(m_e c^2))^2$, where $\mu_e^{1\rightarrow2}$ is defined by Eq.~(\ref{eq:thres-mue}), 
electrons are confined to the lowest Rabi level thus leading to an essentially linear increase of $\bar n^{\rm max}$ with $B_\star$ 
(small nonlinearities are introduced by the electrostatic interactions). Since the electron Fermi energy increases with density, the magnetic 
field is strongly quantising in any layer of the outer crust if $B_\star > B_\star^{\rm drip}\approx 1/2 (\mu_e^{\rm drip}/(m_e c^2))^2$, where 
$\mu_e^{\rm drip}$ is the electron Fermi energy at the neutron-drip transition defined by Eq.~(\ref{eq:muebetan}). Typical values for 
$B_\star^{\rm drip}$ are around 1300~\cite{chapav12,chamel2017b,chasto15,chamut16}. Depending on the value of the magnetic field strength, 
some nuclides may disappear and others appear. For example, the nickel isotopes $^{66}$Ni and $^{64}$Ni are no longer present in the crust 
for $B_\star>67$ and $B_\star>1668$ respectively, whereas $^{88}$Sr and $^{132}$Sn appear at $B_\star=859$ and $B_\star=1989$ respectively (see, e.g. 
Ref.~\cite{chamel2017b} for the detailed composition). All in all, the crust of a neutron star becomes less neutron-rich in the presence of a 
high magnetic field. Moreover, the neutron-drip transition is shifted to either higher or lower densities depending on the magnetic field strength. The 
lowest density is reached for $B_\star=B_\star^{\rm drip}$ and is given by $\rho_{\rm drip}^{\rm min}\approx (3/4) \rho_{\rm drip}(B_\star=0)$
~\cite{chasto15}. 
In the strongly quantising regime, the neutron-drip density increases almost linearly with $B_\star$. 
In all these calculations, the same nuclear masses as in the absence of magnetic fields were employed. However, high 
enough magnetic fields can also influence the structure of nuclei~\cite{pen11,stein16}, inducing additional changes in the crustal composition~\cite{bas15}. 
However, complete nuclear mass tables corrected for the presence of a magnetic field are not yet available.

\subsection{Accreted neutron stars}
\label{sec:outer-crust-accreted}

The composition of the outer crust of an accreting\index{accretion} neutron star in a low-mass X-ray binary can be very different from that of a nonaccreting 
neutron star depending on the duration of accretion. In particular, the original outer crust, containing a mass $\sim 10^{-5}\,M_\odot$ (with $M_\odot$ 
the mass of the Sun), is pushed down by the accreted material and is molten into the liquid core in about $10^4$~years assuming an accretion rate of 
$10^{-9}~M_\odot$ per year. 

The constitution of the outer crust of an accreted neutron star depends on the composition of the X-ray burst ashes, produced in the outermost  
region of the star at densities $\rho \lesssim  10^{7}$~g~cm$^{-3}$. As this material sinks into deeper layers due to accretion, it may undergo 
electroweak and nuclear reactions. The evolution of a matter element was followed in Refs.~\cite{gupta2007,gupta2008,schatz2014,lau2018} considering a reaction network of many nuclei. Such calculations are computationally very expensive, and require the knowledge of a very large number 
of nuclear inputs (e.g. nuclear masses, reaction rates), some of which have not been experimentally measured and must therefore be estimated using 
nuclear models. For this reason, the composition of accreted crusts remains more uncertain than that of catalysed crusts despite recent progress in 
nuclear mass measurements~\cite{estrade2011,meisel2015,meisel2016}. A numerically more tractable approach consists of 
assuming that the rate of an energetically allowed reaction is much faster than the accretion rate. At densities $\rho>10^8$~g~cm$^{-3}$, 
matter is strongly degenerate, and is ``relatively cold''  ($T \lesssim 10^8~{\rm K}$), so that thermonuclear processes are strongly suppressed. 
Under these conditions, the most important reactions are single electron captures\index{electron capture}. Multiple electron captures are very rare and can thus be ignored. 
For instance, the double electron capture by $^{56}$Fe occurs on a timescale of about $10^{20}$~yr~\cite{blaes1990}. 
The first electron capture by a nucleus ($A,Z$) 
\begin{equation}
(A,Z)+e^-\longrightarrow  (A,Z-1)+\nu_e \, ,
\label{eq.sect.accretion.processes.ecap1}
\end{equation}
proceeds in \emph{quasi-equilibrium}: this reaction occurs as soon as the electron Fermi energy $\mu_e$ exceeds some threshold value $\mu_e^\beta$, as determined 
by Eqs.~(\ref{eq:neutron-drip}) and (\ref{eq:muebetan}) with $\Delta Z=1$ and $\Delta N=0$ (in the case of a transition to an excited state of the 
daughter nucleus with energy $E_{\rm ex}$, the mass $M^\prime(A,Z-1)$ must be replaced by $M^\prime(A,Z-1)+E_{\rm ex}/c^2$). In other words, this 
reaction is allowed if the Gibbs free energy per nucleon is lowered. 
The daughter nucleus is generally highly unstable, and captures a second electron \emph{off-equilibrium} with an energy release $Q$: 
\begin{equation}
(A,Z-1)+e^-\longrightarrow  (A,Z-2)+\nu_e + Q \, .
\label{eq.sect.accretion.processes.ecap2}
\end{equation}
With these assumptions, the final composition of accreted neutron-star crusts is independent of the details of the reaction rates. It is only determined by 
the initial composition of X-ray bursts\index{X-ray bursts} ashes and by nuclear masses (as well as the energies $E_{\rm ex}$ for transitions to excited states). Such 
calculations have been carried out in Ref.~\cite{steiner2012} using a liquid droplet model with empirical nuclear-shell corrections. A further simplification 
consists of approximating the distribution of nuclides by a single  nucleus~\cite{haensel1990,haensel2003,haensel2008}. This computationally very fast 
treatment was shown to provide a fairly accurate estimate for the total heat released as compared to reaction network calculations~\cite{haensel2008}. 
The composition of the outer crust of accreted neutron stars have been recently determined using microscopic nuclear mass models~\cite{fantina2018}. 

As the X-ray burst ashes sink into the crust, their proton number decreases due to electron captures whereas their mass number remains unchanged. At some point, 
the daughter nuclei will be so neutron rich that free neutrons will be emitted. This transition, which marks the boundary between the outer and inner regions of the 
crust, will occur when the threshold electron chemical potential $\mu_e^{\rm drip}$  for neutron emission (i.e. $\Delta N>0$ and $\Delta Z=1$) will become lower than 
the threshold electron chemical potential $\mu_e^\beta$ for electron capture alone. This condition can be equivalently expressed as~\cite{cfzh15} 
\begin{equation}
M^\prime(A-\Delta N,Z-1)-M^\prime(A,Z-1)+\Delta N m_n < 0\, .
\end{equation}
Depending on the composition of ashes, and employing microscopic nuclear mass models, the neutron-drip density and pressure are expected to lie in 
the range $\rho_{\rm drip}\sim 2.6-6.5\times 10^{11}$ g~cm$^{-3}$, and $P_{\rm drip}\sim 4.4-13\times 10^{29}$ dyn~cm$^{-2}$ respectively~\cite{fantina2016}.

\section{Inner crust}
\label{sec:inner-crust}
\index{inner crust}

The neutron-saturated clusters constituting the inner crust of a neutron star owe their existence to the presence of a highly degenerate 
surrounding neutron liquid: neutron emission processes, which would lead to the immediate disintegration of these clusters in vacuum, 
are energetically forbidden in a neutron star due to the Pauli exclusion principle since neutron continuum states are already occupied. 
As a newly formed neutron star cools down, free neutrons in the inner crust are expected to become superfluid by forming Cooper pairs 
analogously to electrons in conventional superconductors\footnote{The high temperatures $\sim 10^7$~K prevailing in neutron stars 
prevent the formation of electron pairs recalling that the highest critical temperatures of terrestrial superconductors do not exceed 
$\sim 200$~K~\cite{droz2015}. See also Ref.~\cite{ginzburg1969}.} (see, e.g. Refs.~\cite{lrr,sedrakian2006,margueron2012,page2014,graber2017,chamel2017s} 
for recent reviews; see also Sedrakian\&Haskell in this volume). 

For all these reasons, the inner crust of a neutron star is a unique environment, whose extreme conditions imposed by the huge gravitational 
field of the star cannot be reproduced in terrestrial laboratories. The description of the inner crust of a neutron star must therefore 
rely on theoretical models. Because clusters are inseparable from their neutron environment, both should be consistently treated. Despite 
considerable progress in the development of many-body methods for the description of light and medium mass nuclei on the one hand, and 
infinite homogeneous neutron matter on the other hand (see, e.g. Refs.~\cite{hagen2016b,dickhoff2004,baldo2012,carlson2015,hagen2016,holt2016,herbert2016,oertel2017}), 
ab initio calculations of the inner crust of a neutron star still remain out of reach. Various phenomenological approaches have thus been 
followed (see, e.g. Refs.~\cite{hae07,lrr} for comprehensive reviews), and will be briefly discussed below.

\subsection{Nonaccreted neutron stars}
\label{sec:inner-crust-nonaccreted}
\index{catalysed matter}

State-of-the-art calculations pioneered by Negele and Vautherin in 1973~\cite{nv73} rely on the self-consistent nuclear energy density functional 
theory\index{energy-density functional}, traditionally formulated in terms of effective nucleon-nucleon interactions in the ``mean-field approximation'' (see, e.g. Ref.~\cite{duguet14} 
for a recent review). It should be stressed that in principle, this theory could predict the \emph{exact} ground state of any nuclear system. In practice, 
however, this theory remains phenomenological because the exact nuclear energy density functional is not known. Still, the accuracy of any given functional 
can be tested against experimental nuclear data, as well as results from microscopic calculations following different many-body approaches. The most accurate 
existing functionals are able to fit essentially all measured nuclear masses and charge radii with a root-mean square deviation of about $0.5-0.6$~MeV/$c^2$ 
and $0.03$~fm respectively, 
while reproducing at the same time properties of homogeneous infinite nuclear matter (equation of state, pairing gaps, effective masses) obtained from ab initio 
calculations~\cite{pearson2013,goriely2016}. In the nuclear energy density functional theory, the ground-state energy is obtained by solving the 
self-consistent Hartree-Fock-Bogoliubov (HFB) equations describing independent quasiparticles in an average field induced by the underlying particles
(see, e.g. Ref.~\cite{chamel2013}). In the absence of nuclear pairing, the HFB equations reduce to the 
Hartree-Fock (HF) equations. As recently shown in Ref.~\cite{pastore2017a}, the HFB energy can be estimated with an error of a few keV per particle by 
applying the decoupling approximation, according to which the nuclear pairing phenomenon is described by the Bardeen-Cooper-Schrieffer (BCS) equations~\cite{bcs57}. 
A relativistic version of the nuclear energy density functional theory has been developed, whereby nucleons interact with various effective meson fields (see, e.g., Ref.~\cite{bender2003} for a discussion of both relativistic and nonrelativistic approaches). 
This theory has been usually referred to as relativistic ``mean-field'' (RMF) model\index{relativistic mean-field (RMF) theory}, although it can account for many-body correlations beyond the mean-field 
approximation as its nonrelativistic version. In this approach, causality is guaranteed at high density. It has thus been widely employed in studies of dense-matter in the core of neutron stars~\cite{glen2000}.

Nuclear-energy density functional calculations of neutron-star crusts (see, e.g. Ref.~\cite{margueron2012}) have been traditionally performed using an approximation 
originally introduced by Eugene Wigner and Frederick Seitz in 1933 in solid-state physics~\cite{wigner1933}, whereby clusters with their surrounding 
neutrons are confined inside spherical cells independent from each other. The Wigner-Seitz approximation allows for relatively fast numerical computations. 
The crustal composition obtained in this way was found to be very sensitive to nuclear pairing effects~\cite{baldo2007,grill2011}, but also to the choice of boundary 
conditions~\cite{baldo2007,baldo2006}. This stems from the artificial quantisation of unbound neutron states~\cite{chamel2007,margueron2008,pastore2017b}. 
Over the past years, a few three-dimensional calculations of 
the ground state of cold dense matter have been undertaken in cubic cells with strictly periodic boundary conditions~\cite{magierski2002,gogelein2007}. However, 
the reliability of such calculations is still limited by the appearance of spurious neutron shell effects (see, e.g. Refs.~\cite{newton2009,fattoyev2017}) that can 
only be completely alleviated either by choosing a large enough cell (much larger than the screening length)~\cite{sebille2009,sebille2011} or by considering a single Wigner-cell of the crystal lattice 
(a truncated octahedron in case of a body-centred cubic lattice) with Bloch boundary conditions~\cite{chamel2012,schuetrumpf2015}. In the former case, the results 
could also be biased by the arbitrary choice of the shape of the ``supercell'' (see, e.g., Ref.~\cite{gimenez2015}). 
The latter approach is computationally more tractable due to the reduced size of the cell and the smaller number of particles. On the other hand, it requires the prior  
knowledge of the crystal structure. This, however, does not appear as a serious limitation, except for describing the densest part of the crust 
(see Section~\ref{sec:pastas}). Indeed, clusters are generally assumed to form a body-centred cubic lattice (in some layers, however, a face-centred cubic lattice
might be energetically favoured, as found, e.g. in Ref.~\cite{oka13}). Although the presence of the neutron liquid 
leads to an induced interaction between clusters that could potentially affect the structure of the inner crust~\cite{kobyakov2014}, refined estimates of the induced 
interaction have dismissed this possibility~\cite{kobyakov2016}. Whether the crust is described by a supercell with periodic boundary conditions or by 
a single Wigner-Seitz cell with Bloch boundary conditions, nuclear-energy density functional calculations are computationally extremely costly. 

For this reason, many studies of the inner crust of a neutron star rely on compressible liquid drop models (see, e.g. 
Refs.~\cite{douchin2001,newton2013,deibel2014,gulminelli2015,bao2016,fortin2016,tews2017,lim2017} for recent calculations). In this approach, the nuclear clusters 
and the neutron liquid are considered as two distinct homogeneous phases. The bulk and surface properties of these two phases can in principle 
be determined using the nuclear energy density functional theory. However, empirical parametrisations of the surface properties are often employed. 
Despite its simplicity, the liquid drop model has shed light on various aspects of dense matter. In particular, the formation of neutron-proton clusters 
was shown to arise from a detailed balance between Coulomb and surface effects~\cite{BBP1971}, leading to the prediction of complex configurations commonly 
referred to as ``pastas'' in the densest layers of the crust 
(see Section~\ref{sec:pastas}). Moreover, the liquid drop model has 
been employed to study the role of nuclear parameters, such as the symmetry energy, on the crustal composition and the crust-core boundary (see, 
e.g., Refs.~\cite{newton2013,bao2016}).

A more realistic treatment of dense matter in neutron-star crusts is to employ semi-classical methods such as the Thomas-Fermi (TF) approximation (see, e.g. 
Refs.~\cite{bcpm,gogelein2007,oka13,fortin2016,lim2017,oya07,avancini08,miyatsu2013,grill14,iida2014}), or its higher-order extensions ~\cite{onsi1997,gor2008,mu15}. The neutron liquid and the clusters 
are not treated separately, but are described in terms of continuous neutron and proton distributions. This method is based on a systematic expansion of the smooth 
part of the single-particle quantum density of states in powers of $\hbar$ (see, e.g., Ref.~\cite{brack1985} for a review; see Ref.~\cite{centelles1993} for the 
relativistic version). The accuracy of this development has been recently studied in Refs.~\cite{papa13,aymard14}.
The shell correction to the total energy 
of a Wigner-Seitz cell (responsible for the oscillatory part of the single-particle quantum density of states) can be added perturbatively via the Strutinsky 
integral~\cite{dutta2004,onsi2008,pearson2012,pearson2015,pearson2018}. This so called ETFSI method is not only a computationally high-speed approximation to the full 
HF+BCS equations, but it also avoids the pitfalls discussed above that plague the current numerical implementations of the nuclear energy density functional theory. 
These calculations show that proton-shell effects still play an important role in determining the ground-state composition of the inner crust, with clusters 
containing predominantly $Z=40$ protons~\cite{pearson2012}. The appearance of this magic number can be understood from the fact that the strength of the 
spin-orbit coupling tends to decrease with the neutron excess, as measured in ordinary nuclei~\cite{schiffer2004}. However, the importance of shell effects is 
mitigated by pairing~\cite{pearson2015}. Depending on the symmetry energy, clusters with $Z=20, 50, 58$ or $92$ can also be favored~\cite{pearson2018}. In the densest region of the crust, configurations with different cluster sizes may differ by a few keV per nucleon at 
most, suggesting that this region could be very heterogeneous. Further studies beyond the mean-nucleus approximation should thus be pursued. At high enough density, protons may drip out of clusters (see, e.g. Refs.~\cite{pearson2018,pethick1995}), and are expected to become superconducting at low temperatures. 

With further compression, the crust dissolves into an homogeneous mixture of nucleons and electrons. The crust-core transition appears to be essentially continuous from the thermodynamic point of view, and occurs at a density between about one third and one half of the nuclear saturation density depending on the symmetry energy and on the nuclear model employed (see, e.g. Refs.~\cite{newton2013,bao2016,ducoin2011,iida2014,sulaksono2014,seif2014,boquera2017,fang2017b} for recent studies). 

\subsection{Highly magnetised neutron stars}
\label{sec:inner-crust-magnetar}

As discussed in Sections~\ref{sec:surface-magnetars} and \ref{sec:outer-crust-magnetar}, the presence of a high magnetic field may have a profound influence on 
the composition and on the properties of dense matter. However, few studies have been devoted so far to the role of a high magnetic field on the inner crust of 
a neutron star. Rabi quantisation of electron motions was studied in Refs.~\cite{nandi2011,nandi2013,nandi2011b} within the TF approximation. It was found that
the inner crust of highly magnetised neutron stars\index{magnetar} is more proton rich, similarly to the results obtained in the outer crust (see 
Section~\ref{sec:outer-crust-magnetar}). As a consequence, magnetised crusts contain larger clusters separated by a smaller distance, whereas the neutron 
liquid is more dilute. However, these effects were found to be negligible for magnetic field strengths below $\sim 10^{16}-10^{17}$~G. In addition to the modification 
of the electron motion, the effects of the magnetic field on nucleons was studied in Ref.~\cite{lima2013} within the TF approximation. The transition density 
between different cluster types was found to exhibit typical oscillations as a function of the magnetic field strength, as also found in the outer crust
for the density at the interface between different layers. Studies of spatial instabilities in highly magnetised homogeneous stellar matter within the RMF approach 
and taking into account the anomalous magnetic moment of nucleons suggest that the densest part of the crust could consist of alternating layers of 
homogeneous and inhomogeneous regions~\cite{fang2016,fang2017,chen2017}. However, quantisation effects of the magnetic field on 
the crust-core transition are washed out at temperatures of order $10^9$~K~\cite{fang2017b}. The spin polarisation of matter adds to the complexity of 
these phases (see, e.g. Ref.~\cite{rabhi2015} and references therein). Moreover, the formation of Cooper pairs in the spin-singlet channel thus becomes 
disfavoured in the presence a high magnetic field. Calculations in pure neutron matter suggest that neutron superfluidity in the crust of a neutron star is 
destroyed if the magnetic field strength exceeds $10^{17}$~G~\cite{stein2016}. However, calculations of inhomogeneous neutron superfluidity in magnetised crusts are still lacking. This warrants further studies.

\subsection{Accreted neutron stars}
\label{sec:inner-crust-accreted}

Comparatively very few studies have been devoted to the inner crust of accreted\index{accretion} neutron stars. This stems from the fact that the rates of electroweak and 
strong nuclear reactions that could possibly occur in this environment where  neutron-proton clusters coexist with free neutrons are poorly known. Moreover, 
a very large number of different nuclear species are expected to exist in any given layer of the accreted crust. Because these clusters owe their existence to the 
surrounding neutrons, their properties cannot be experimentally measured. A consistent treatment of a distribution of clusters with free neutrons in the framework 
of the nuclear energy-density functional theory is computationally extremely challenging as this would require solving the HFB equations in considerably larger 
cells than in nonaccreted crusts.  

For all these reasons, the composition of the inner crust of accreted neutron stars has been mainly studied so far using compressible liquid-drop models, in the single-nucleus 
approximation (see, Refs.~\cite{haensel1990,haensel2003,haensel2008} and references therein) and taking into account the distribution of different 
clusters~\cite{steiner2012}. The composition was determined by following the sequence of electron captures, neutron emissions, and pycnonuclear reactions that occur 
as the elements present at the bottom of the outer crust sink into deeper layers. These calculations indicate that the accretion of matter onto a neutron star can radically change the internal constitution of the inner crust. Very recently, the role of nuclear shell effects on the properties of accreted crusts has been studied in the framework of the nuclear energy density functional theory using the ETFSI method~\cite{fantina2018}.

\section{Nuclear pasta mantle}
\label{sec:pastas}

In the outer crust of a neutron star, nucleons are bound inside nuclei, whose size (typically 5-6 fm) is negligibly small compared to the 
ion spacing ($a_{\rm N}$ varies from about $10^5$ fm at the surface to $10^2$ fm at the neutron-drip point). Consequently, the composition 
of the outer crust (determined by nuclear physics) and its structure (determined by the physics of dense Coulomb plasmas) can be studied 
separately. Whereas the spatial arrangement of nuclei in the outer crust is governed by the long-range Coulomb interactions, the shape and 
size of an individual nucleus are controlled by the short-range nuclear and Coulomb interactions. Nuclei expected to be present in the 
outer crust of a neutron star are almost spherical. Considering axially symmetric nuclei, the nuclear surface can be parametrised by the 
radius $R_N(\theta)=R_0(1+\beta_2 Y_2^0(\theta)+\beta_4 Y_4^0(\theta))$, where $\theta$ is the polar angle, $Y_\ell^m$ are spherical harmonics, 
$\beta_2$ and $\beta_4$ are dimensionless parameters characterising quadrupole and hexadecapole deformations. Positive (respectively negative) 
values for $\beta_2$ lead to prolate (respectively oblate) spheroid nuclei. Typical orders of magnitude for these latter parameters are 
$|\beta_2| \sim 10^{-1}$ and $|\beta_4|\sim 10^{-2}$. 
%
%
In most regions of the inner crust, nuclear clusters remain sufficiently far apart that their structure is not influenced by the presence of 
neighbouring clusters. Therefore, clusters are expected to be quasi spherical as in the outer crust. However, because the average distance between clusters 
decreases with increasing density, clusters eventually fuse and connect into herringbone structures~\cite{watanabe2009} similarly to percolating 
networks as speculated earlier by Ogasawara \& Sato~\cite{ogasawara1982}. 

Matter in the densest layers of the crust is frustrated and various 
exotic configurations collectively referred to as ``nuclear pastas''\index{pasta phases} might appear, as first shown in Refs.~\cite{ravenhall1983,hashimoto1984,oyamatsu1984} 
(see, e.g., Refs.~\cite{lrr,watanabe2012} for a review). Frustrated nuclear systems are also encountered in heavy-ion collisions although under different physical conditions (see, e.g. Ref.~\cite{botvina2005}). 
If they exist, nuclear pastas would behave like liquid crystals and would thus form a liquid mantle below the crust~\cite{pethick1998}. Nuclear pastas 
have been studied following the same treatments as described in Section~\ref{sec:inner-crust} for the inner crust: compressible liquid drop models (see, e.g. 
Refs.~\cite{newton2013,lim2017,nakazato2011,gupta2013,vinas2017} for recent calculations), semi-classical methods (see, e.g. Refs.~\cite{bcpm,gogelein2007,oka13,
fortin2016,lim2017,oya07,avancini08,grill14,iida2014,mu15,vinas2017}, and the nuclear energy density functional theory (see, e.g. Refs.~\cite{magierski2002,gogelein2007,
newton2009,fattoyev2017,sebille2009,sebille2011,schuetrumpf2015,sagert2016,pais2012}). The existence of nuclear pastas have been also investigated using the quark-meson 
coupling model~\cite{grams2017}, in which nucleons in dense matter are described by non-overlapping static spherical ``bags'' of quarks interacting through the 
self-consistent exchange of mesons in the mean-field approximation. In all these approaches, a few specific nuclear shapes are generally considered, or some 
symmetries are assumed. 

The formation of nuclear pastas has been explored using molecular dynamics (see, e.g. Ref.~\cite{dorso2012b} for a review). By treating 
nucleons as classical pointlike particles interacting through a two-body potential, classical molecular dynamics allow for large-scale simulations with 
$\sim 10^3-10^5$ particles in a box with periodic boundary conditions~\cite{dorso2012,schneider2013,schneider2014,horowitz2015,schneider2016,berry2016}. Although 
no other constraints are imposed, the nuclear pasta configurations obtained from these simulations could still be influenced by the geometry of the box~\cite{gimenez2015} 
and the treatment of Coulomb interactions~\cite{alcain2014}. Quantum molecular dynamics simulations (see, e.g. Refs.~\cite{watanabe2012,maruyama2012} for a review) 
rely on a semi-classical description in which the state of a nucleon is represented by a Gaussian wave packet moving (classically) in a mean field. The antisymmetrisation 
of the many-body wave function is implemented in an effective way through a phenomenological repulsive potential, so called Pauli potential. Calculations have been 
performed with $\sim 10^3$ nucleons~\cite{watanabe2012,maruyama2012,nandi2016}. Fermi statistics is properly taken into account in fermionic molecular dynamics, in 
which the many-body wavefunction is expressed as a Slater determinant. However, such simulations are computationally extremely costly (the computing time scales as 
$N^4$ as compared to $N^2$ for classical and quantum molecular dynamics, where $N$ is the number of particles). For this reason, only a few fermionic molecular 
dynamics simulations have been carried out so far~\cite{vantournhout2011,vantournhout2012}. Finally, nuclear pastas have been studied using a nuclear analog of the 
Ising model, in which the nucleon positions are discretized in a three-dimensional cubic lattice, and assuming that each lattice site can be occupied by only one 
nucleon at most~\cite{hasnaoui2013}. 

The presence of nuclear pastas in neutron stars remains uncertain and model-dependent. Still, all models supporting their existence predict the following sequence 
of configurations with increasing density: polpette/gnocchi (spherical clusters), spaghetti (cylindrical clusters), lasagne (slabs), bucatini/penne/maccheroni (cylindrical holes), 
Swiss cheese (spherical holes). Intermediate phases have been found by some models such as ``nuclear waffles'' (slabs with holes)~\cite{sebille2011,schneider2014,williams1985}, 
cross-rods~\cite{pais2012,schneider2014,lassaut1987}, or slabs connected by helical ramps~\cite{berry2016}. Because the nuclear pasta mantle is very dense, 
it may represent a sizable fraction of the crustal mass~\cite{lorenz1993}, and thus may have important astrophysical consequences (see, e.g., Ref.~\cite{lrr}). 

As recently discussed in Ref. \cite{fortin2016} (see also the contribution from Fantina and Burgio in this volume), a thermodynamical inconsistent treatment of the crust-core boundary can lead to significant errors on the predicted masses and radii of neutron stars. Moreover, thermodynamical inconsistencies could trigger spurious hydrodynamical instabilities. A unified description based on the same model is therefore of utmost importance. We shall come back to this issue when addressing the quark-hadron phase transition in the core.

\section{Nuclear and hypernuclear matter}

\begin{figure}
	\includegraphics[width=0.8\textwidth]{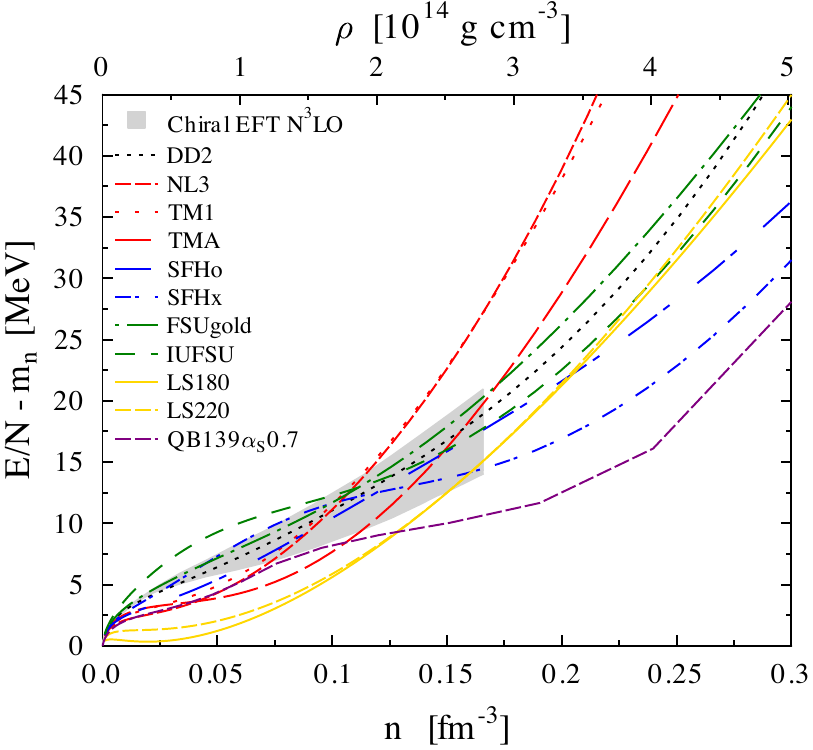}
	\caption{Energy per baryon in pure neutron matter for different supernova EoS, compared to results of
	$\chi$EFT (grey band \cite{Kruger:2013kua}), from Ref.~\cite{Fischer:2013eka}. }
	\label{fig:ChEFT}
\end{figure}

A recent review about the equation of state (EoS) for supernovae, neutron stars and their mergers has been given in 
\cite{Oertel:2016bki}, where also the very different theoretical approaches to the nuclear matter problem are shortly reviewed (see also Fantina and Burgio in this volume). 
Very roughly one can distinguish two classes of systematic approaches to nuclear and 
hypernuclear matter\index{hypernuclear matter}: the ones based on (realistic) baryon-baryon interactions that can describe also the phase shift data as well as properties of nuclear and hypernuclear clusters on the one hand and the approaches based on relativistic density functionals with elaborated (density-dependent) couplings to mesonic mean fields that are fitted to describe observable properties of large nuclei and predict the behaviour of the EoS at saturation densities and beyond. 
While the nonrelativistic approaches based on potentials (Brueckner-Hartree-Fock\index{Brueckner-Hartree-Fock}, Brueckner-Bethe-Goldstone\index{Brueckner-Bethe-Goldstone}, Variational approach) require three-body forces in order to obtain acceptable ground state properties of nuclear matter and the $2M_\odot$ constraint, the relativistic Dirac-Brueckner-Hartree-Fock (DBHF)\index{Dirac-Brueckner-Hartree-Fock (DBHF)} approach fulfills constraints from compact star physics and heavy-ion collisions without them \cite{Klahn:2006ir}.  
A modern constraint for theories of nuclear matter is provided by the ab-initio chiral effective field theory ($\chi$EFT)\index{chiral effective field theory ($\chi$EFT)} approach that provides a link between QCD (in its low-energy limit as chiral perturbation theory) and nuclear matter properties with nucleons, pions and a nucleon-nucleon contact interaction as the elements of a systematic perturbation theory. The $\chi$EFT shall be considered as a benchmark for the pure neutron matter case (no bound states) at subnuclear densities where errors due to unaccounted higher orders are sufficiently small. See Fig.~\ref{fig:ChEFT} for a comparison of approaches. 

We would like to note that the density-dependent relativistic mean-field theory\index{relativistic mean-field (RMF) theory} "DD2" lies perfectly within the bounds provided by the grey band of the $\chi$EFT results. 
Unfortunately, this band broadens quickly with density so that at supersaturation densities it does no longer provide reliable constraints. 

In order to explore the nuclear EoS beyond the saturation density of $n_0=0.15$ fm$^{-3}$, it has therefore been suggested \cite{Hebeler:2013nza,Read:2008iy,Raithel:2016bux} to consider sequences of polytropic EoS \index{polytropic EoS}
\begin{equation}
\label{polytrope}
P(n) =  \kappa_i  (n/n_0)^{\Gamma_i}, \  n_i < n < n_{i+1}, \ i=1 \dots N,
\end{equation}
and by varying the polytropic indices $\Gamma_i$ (and eventually also the switch densities $n_i$) to obtain a most generic class of high-density EoS from which the one(s) can be chosen that perform best with respect to constraints from compact star observations. 
Such a multi-polytropic ansatz is also capable of describing a hadron-to-quark matter phase transition, with a constraint pressure region (as in the Maxwell construction case) or with changing pressure as for the case of an extended mixed (pasta) phase in the compact star \cite{Alvarez-Castillo:2017qki,Alvarez-Castillo:2017xvu,Zdunik:2005kh}.   
We shall come back to this in Sect.~\ref{sec:HSP} below.

\subsection{Hyperons and hyperon puzzle}

At densities of two to three times saturation density the hyperon threshold is expected to be
crossed. 
It has been shown within Brueckner-Bethe-Goldstone\index{Brueckner-Bethe-Goldstone} calculations \cite{Baldo:1999rq}
that under conservative assumptions 
for the forces involving hyperons the appearance of these additional degrees of freedom softens the EoS and hence lowers the maximum mass of a neutron star so that even the well-constrained binary radio pulsar masses of typically $\sim1.4~M_\odot$ cannot be reached.
Without additional repulsive hyperon interactions this is exactly what all neutron star calculations based 
on nuclear hyperon model EoSs predict.
This is called the "hyperon puzzle" \index{hyperon puzzle}\cite{Zdunik:2012dj}.
In general, the EoS beyond saturation density is not well constrained~\cite{Oertel:2016bki}.
For hyperons and their interaction with nucleons and themselves the situation is even worse.
However, from the observed existence of hyperons 
and massive neutron stars 
it seems evident that nucleon-hyperon and likely hyperon-hyperon repulsion \cite{Rijken:2016uon}
plays an important role
(unless hyperons themselves play no decisive role for neutron star structure).
Meanwhile, different approaches have been successfully applied which 
allow for stable neutron stars with up to two solar masses.
Repulsion stiffens the hyperon matter EoS 
sufficiently to account for massive neutron stars. 
At the same time the stiffening results in
an onset of hyperon degrees of freedom at higher densities.
Thus, the fraction of hyperons is reduced.
In another scenario this idea is taken to the limit where the hyperon onset density is
larger than the densities one would find in a neutron star.
Generally, in RMF models for hypernuclear matter \index{hypernuclear matter} \cite{Hofmann:2000mc}
there is no severe hyperon puzzle, 
because the unknown density dependent scaling of the 
meson masses and couplings of the can be defined in the spirit of a relativistic density functional theory such that all known constraints on the stiffness of the EoS from heavy-ion collisions and compact star observations can be fulfilled 
\cite{Maslov:2015msa,Maslov:2015wba}.
A particular role plays the sufficiently repulsive effect of the $\phi-$meson mean field
\cite{Weissenborn:2011ut}. 
This solution of the hyperon puzzle holds also for hypermatter stars with (color superconducting) quark matter cores \cite{Bonanno:2011ch,Lastowiecki:2011hh}.
A third solution to the hyperon problem is a transition to sufficiently stiff quark matter \cite{Baldo:2003vx}
which can happen before or after the transition to hyperon matter.

\subsection{$\Delta$ isobars and $\Delta$ puzzle}
\label{ssec:delta}
While a large amount of literature is devoted to the discussion of hyperons and the hyperon puzzle,
only little work has been done to discuss the presence of $\Delta(1232)$ isobars 
\index{$\Delta(1232)$ isobars} which, once they appear in neutron star matter, have a similar softening effect on the EoS as the hyperons. 
That $\Delta$ isobars have so far been largely neglected may be due to the outcome of the seminal paper \cite{Glendenning:1984jr} which concluded that they may appear only at too high densities to be relevant for the structure of compact star interiors.
In Refs.~\cite{Drago:2013fsa,Drago:2014oja} it has been shown that the threshold density strictly correlates with the $L$ parameter, the  derivative of the symmetry energy in nuclear matter, which according to the compilation of constraints in 
\cite{Lattimer:2012xj}
lies in the range of $40.5 {\rm MeV} \le L \le 61.9$ MeV.
For the relativistic mean field EoS SFHo \cite{Steiner:2012rk}
in this range the $\Delta^-$ isobar occurs even at lower densities than the $\Lambda$ hyperon in neutron star matter and leads to a further softening of the EoS. Thus the $\Delta$ isobars too limit the maximum mass of a hadronic compact star to $\sim 1.5 - 1.6$ M$_\odot$ 
\cite{Drago:2014oja}
and entail a $\Delta$ puzzle \index{$\Delta$ puzzle}. 

A solution of this puzzle may be similar in spirit to those possibilities discussed as solutions of the hyperon puzzle, namely, that there could be stiffening effects in the hadronic EoS at supernuclear densities (due to, e.g., multi-pomeron exchange\index{multi-pomeron exchange} \cite{Yamamoto:2015lwa}
in the baryon-baryon interaction or the density dependence of meson masses and their couplings in RMF models \cite{Kolomeitsev:2016ptu}
)
or a sufficiently early phase transition to stiff quark matter could occur.
As an equilibrium phase transition, the latter possibility would require a rapid stiffening of the quark matter EoS after the transition so that a maximum mass above the 2~M$_\odot$ constraint could be reached. This would entail the "reconfinement" problem\index{reconfinement problem} as in the case of the corresponding solution of the hyperon puzzle. 
In order to circumvent problems with unphysical crossing of EoS an interpolation scheme has been suggested which we discuss more in detail in subsection \ref{ssec:interpol} below.
This scheme realises a crossover transition between two phases which leads to an intermediate phase that is stiffer than those it is bridging. The mass-radius relation obtained with such a construction applied to several soft hadronic EoS shows a sufficiently large maximum mass and an increase in radius at the transition. 

Another solution of the $\Delta$ puzzle has been suggested within the scenario of coexistence of 
{\it two families}\index{two families scenario} of compact stars 
\cite{Berezhiani:2002ks,Drago:2004vu,Bombaci:2004mt},
where the second family may belong to the class of strange quark matter stars with a sufficiently large maximum mass to fulfill the 2~M$_\odot$ constraint.
The hadronic branch of compact star configurations would be metastable\index{metastable hadronic stars} against the "decay" to the strange quark matter branch by a nucleation of strange quark matter droplets in the stars interior.
The nucleation time scales could be sufficiently long to populate the branches of both families of solutions. The compact stars with smaller maximum mass would then be hadronic stars while the more massive stars with larger radii would be strange quark matter stars. For details, see 
\cite{Drago:2013fsa}
The main caveat of the two-family scenario is that the timescale for the nucleation of a strange quark matter droplet in the metastable hadronic phase by quantum tunneling 
\cite{Iida:1998pi,Bombaci:2006cs}
sensitively depends on the value of the bag constant employed in those models (and also on other parameters, if present).
This makes any predictions for the lifetime of the metastable state rather arbitrary. 
Once a strange quark matter droplet is formed, the conversion of the hadronic star to a strange quark star proceeds on a millisecond timescale as has been demonstrated in simulations 
of the turbulent combustion process 
\cite{Herzog:2011sn,Pagliara:2013tza}.

\subsection{The case of the $d^*(2380)$ dibaryon}
The recent discovery of the $d^*(2380)$ dibaryon\index{$d^*(2380)$ dibaryon} (see \cite{Clement:2016vnl}
for a recent review) has triggered an investigation of its possible role in compact star interiors \cite{Vidana:2017qey}. It has been found that the $d^*(2380)$ would appear at densities around three times nuclear saturation density and comprise about $20\%$ of the matter in the core of massive compact stars and even higher fractions possible in the course of compact star merger events. 
The formation of  the $d^*(2380)$ dibaryon in compact star matter can be understood as the conversion of pairs of neutrons and protons, thus effectively reducing their number density and introducing a rather heavy degree of freedom instead. This leads to a strong softening of the equation of state, an effect like the occurrence of a mixed quark-hadron matter phase.
The presence of the dibaryon in the composition of a high-mass neutron star has an effect on the cooling of the star since it adds fast direct Urca type cooling processes, similar to the pair-breaking and pair-formation processes in superfluid neutron star matter.

\subsection{Chirally symmetric nuclear matter phase}

There is evidence from recent finite-temperature lattice QCD simulations of the FASTSUM collaboration that the restoration of chiral symmetry takes place already within the hadronic phase by parity doubling
\index{parity doubling} \cite{Aarts:2017rrl},
which is then signalled by a mass degeneracy of the baryonic chiral partner states (the nucleon and the $N^*(1535)$) in such a way that the lower-mass state is almost independent of the medium while the higher mass state comes down (as in the well-known case of pion and the sigma meson described by the NJL model, see \cite{Klevansky:1992qe}
and references therein).
A theoretical description can be based on the parity doublet model 
\cite{Detar:1988kn,Jido:2000bw,Jido:2001nt}
which has been applied to the phenomenology of hot dense matter, including neutron stars 
\cite{Dexheimer:2007tn,Mukherjee:2017jzi}
A recent extension of the model to the quark level of description can address a deconfinement transition 
\cite{Benic:2015pia}
and has been applied to the discussion of compact star phenomenology 
\cite{Marczenko:2018jui}
In this work it was shown that the account for the quark substructure of the baryons entails a stiffening of the asymmetric nuclear matter\index{stiffness of the EoS} just above the saturation density before the actual chiral restoration occurs which results in the doubling of nucleonic degrees of freedom
(parity doubling). This transition proceeds as a rather strong first-order phase transition that entails an almost horizontal branch at $\sim 2$ M$_\odot$ in the corresponding $M-R$ diagram, still before quark deconfinement. 
A pattern like this had been suggested earlier as the signature of a strong deconfinement transition \cite{Alvarez-Castillo:2016wqj}
In view of this possible strong effect of the combination of quark substructure and parity doubling on the hadronic EoS this suggestion should be revisited. 
Another prediction of the parity doubling model is a new formula for the threshold value of the proton fraction above which the fast direct URCA\index{URCA process} cooling process would become operative in compact stars. 
Thus the chiral symmetry restoration by parity doubling in the nuclear EoS should be taken into the consideration of modern compact star cooling theories that are employed when interpreting cooling data. 
 
\subsection{Stiffness from quark Pauli blocking\index{quark Pauli blocking}}

We have just discussed the importance of repulsive interactions for solving the hyperon puzzle\index{hyperon puzzle}. 
But how generic is the presence of repulsive interactions in hadron-hadron scattering and in dense hadronic matter and what is their origin? Is it just the exchange of vector mesons ($\omega$, $\phi$, $\varrho$, ...)? 
There is also the concept of the excluded volume, based on the fact that baryons (hadrons in general) are finite size objects due to their quark substructure, that gives rise to an increase in the pressure of the system, like a repulsive interaction. 
For standard excluded volume approaches the pressure even diverges when the available volume goes to zero (closest packing of hard-sphere hadrons). 
How to avoid such a divergence?
In simple terms, a hadron can be seen as an MIT bag of quarks with a radius. 
When at high densities the bags merge the individuality of hadrons gets lost and a big bag filled with quark matter emerges, described by the thermodynamical bag model.   
This solution, the phase transition to quark matter, was already suggested for the Hagedorn resonance gas model by Hagedorn and Rafelski \cite{Hagedorn:1980cv,Rafelski:1980rk}.

Another solution has been suggested as a reformulation of the concept of the excluded volume 
\index{excluded volume} so that the available volume is a function of the density which deviates from the system volume as a function of the density so that below a certain density there is no effect, but for high densities it goes to zero asymptotically, for instance as a Gaussian function. 
For details of the thermodynamically consistent formulation of such an approach, see \cite{Typel:2016srf}.
This concept can be carried even further by allowing the available volume to be a function not only of the density, but also of the temperature and to exceed the system volume so that attractive interactions can be modeled too. Recently, with this generalization of the concept a QCD phase diagram \index{QCD phase diagram} has been obtained that resembles first order and crossover transitions, separated by a critical endpoint \cite{Typel:2017vif}, as expected for full QCD. 

Despite this phenomenological success and its thermodynamic consistency, the excluded volume model does not provide a microphysical understanding for the origin of the repulsion between hadrons and its medium dependence which determines the stiffness of high-density matter. 
A satisfactory explanation can be given by accounting for the Pauli exclusion principle on the quark level which leads to a repulsive quark exchange interaction between hadrons, very similar to molecular systems where the apparent hard sphere behavior of atoms and molecules originates from the Pauli blocking among electrons in  the atomic orbitals. 
A quantitative estimate for the Pauli blocking effect in nuclear matter has been given on the basis of a simple string model for quark confinement and the baryon spectrum in Ref.~\cite{Ropke:1986qs}, where it was found that the result very well corresponds to the repulsive part of the density dependent repulsion in the Skyrme-type model by Vautherin and Brink \cite{Vautherin:1971aw}. 
The Pauli quenching effect is not only density- but also temperature and momentum dependent so that one can deduce from it a temperature and density dependent effective nucleon mass \cite{Ropke:1986mh}.
For the applications to compact star physics it is very important that the Pauli blocking is flavor dependent. It gives a major contribution to the symmetry energy\index{symmetry energy} and allows, in principle, to determine the contribution to the pressure of hypernuclear matter\index{hypernuclear matter}. 

Quark Pauli blocking\index{quark Pauli blocking} leads to a positive energy shift of baryons, so that the occurrence of hyperons and $\Delta$ isobars in dense matter may become inhibited as for $T=0$ matter the critical densities for the occurrence of these species are shifted to higher values.
Pauli blocking is due to an admixture of 6-quark or even higher multiquark wave functions in dense baryonic matter. 
It may on the other hand provide a mechanism for binding two $\Delta$ isobars in the $d^*(2380)$ dibaryon by quark exchange forces since a $\Delta-\Delta$ molecule has overlap with a three-diquark state. 

As in the simple-minded picture of merging bags the quark exchange effect between hadrons is a precursor of quark deconfinement! 
The additional pressure drives the possible phase transition to lower densities, the effect of the antisymmetrisation among quarks of overlapping hadron wave functions generates multiquark components of the many-particle wave function in dense nuclear matter. 
With such a microphysical background an early calculation of stable hybrid stars with interacting quark matter cores at high-mass could be presented \cite{Blaschke:1989nn}. 
The effect of partial quark delocalisation leads to the emergence of a quark Fermi sea. Such effects have been described as typical for quarkyonic matter\index{quarkyonic matter} \cite{Kojo:2009ha,Kojo:2010fe}.
The question arises how partial chiral symmetry restoration in dense nuclear matter could modify the Pauli blocking effect. In a recent exploratory calculation that used the Pauli quenching result of \cite{Ropke:1986qs,Blaschke:1989nn} with density-dependent quark masses, a strong enhancement of the effect has been found with results for the equation of state \cite{Blaschke:2018} that correspond to those from the modified excluded volume model by Typel \cite{Typel:2016srf}.

\section{Quark matter}

\begin{figure}
	\includegraphics[width=0.4\textwidth]{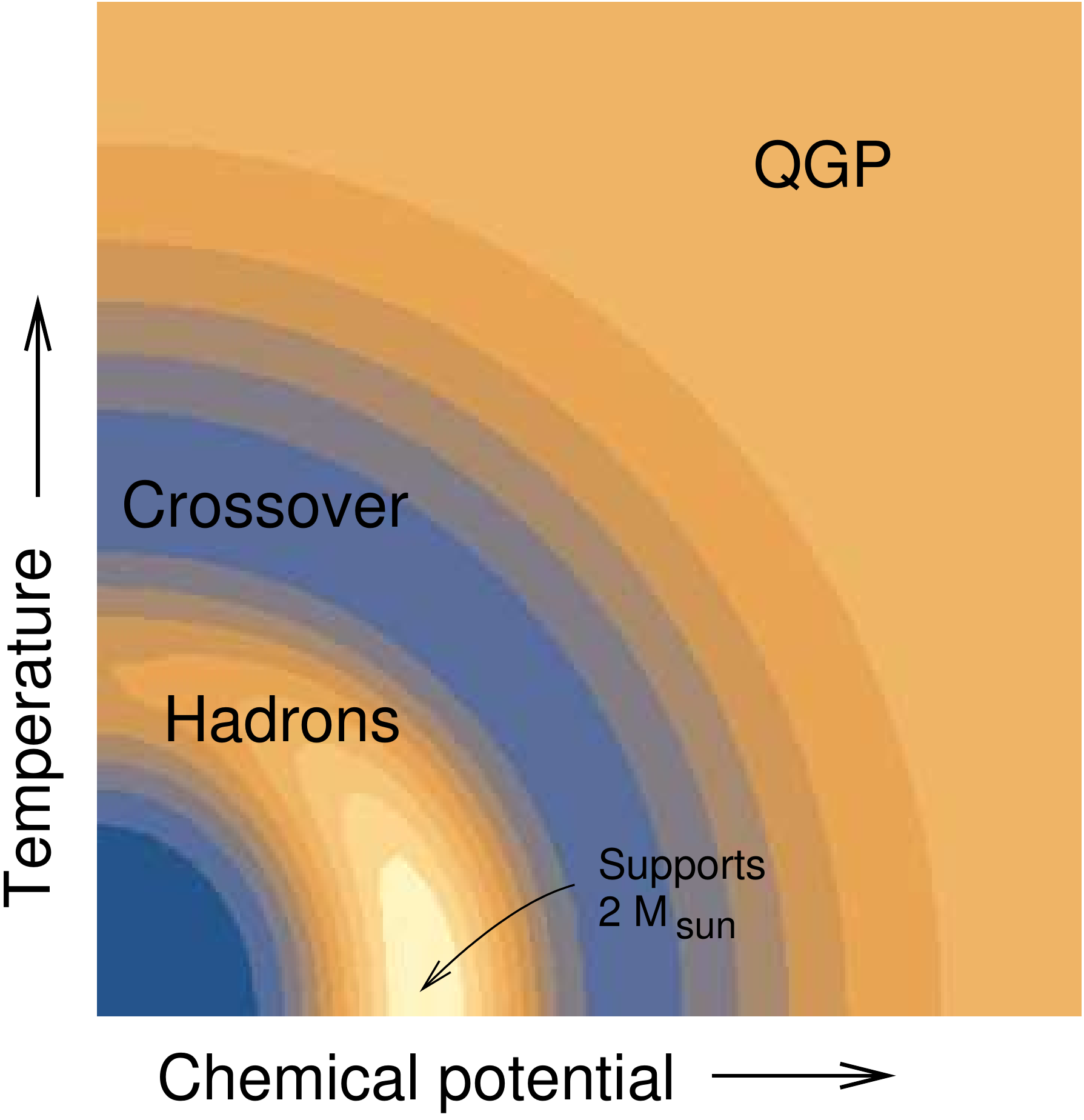}
	\includegraphics[width=0.6\textwidth]{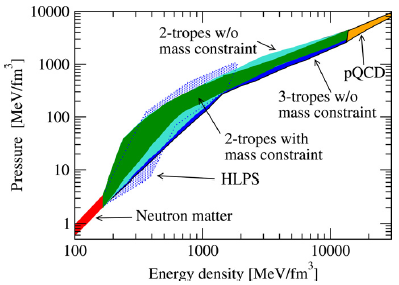}
	\caption{Left panel: Schematic view of the speed of sound in the QCD phase diagram [This figure was kindly provided by M.G. Alford and C.J. Horowitz who created it during the INT-16-2b program on "The Phases of Dense Matter" at the INT Seattle (2016)]. Dark color stands for low $c_s$, for asymptotic values of temperature and chemical potential $c_s^2=1/3$. At low temperatures, in-between nuclear saturation density and the perturbative QCD\index{perturbative QCD} region, $c_s$ must come close to 1 (speed of light), for the EoS to fulfill the $2~M_\odot$ constraint on the pulsar mass.   
		Right panel: Multi-polytrope interpolations of the compact star EoS between the well established limits of pure neutron matter (in red) and perturbative QCD (in yellow), from \cite{Kurkela:2014vha}. 
		Compact star phenomenology can constrain the EoS for energy densities up to about 2~GeV/fm$^3$,
		see \cite{Hebeler:2013nza} (HLPS) and \cite{Alvarez-Castillo:2017qki}.}
\label{fig:eos-constraint}
\end{figure}

Deconfined quark matter\index{deconfined quark matter} shall be described in terms of quarks and gluons, the degrees of freedom of QCD,
the gauge field theory of strong interactions. 
For applications in neutron stars, i.e. at high densities and zero temperature, the ab initio approach to solve QCD numerically by lattice QCD simulations\index{lattice QCD simulations} , is not available due to the sign problem. 
At zero baryon density the lattice QCD simulations have reached a quality where they reproduce at low temperatures the hadron mass spectrum and the hadron resonance gas thermodynamics which can be directly compared with phenomenological approaches. 
It goes so far that up to temperatures including the pseudocritical temperature of the hadron-to-quark matter crossover transition the thermodynamics of QCD can be described as a hadron resonance gas. 
The lesson for the $T=0$ domain of neutron star physics therefore can be that it has to develop effective field theory approaches which are nonperturbative so that they can address the formation of hadronic (baryonic) bound states as well as collective phenomena like dynamical chiral symmetry breaking and color superconductivity. At the same time, the effective models shall try to capture as many aspects of QCD as possible, like the (approximate) chiral symmetry of its Lagrangian in the light quark sector and its dynamical breaking. See Fig.~\ref{fig:eos-constraint}.

In this section we want to recapitulate a few challenging questions in this context and discuss the progress which has been made in these directions, in particular with the help of the COST Action "NewCompStar".

\subsection{Asymptotically high densities}

Although one can hope that perturbative results will constrain effective models at large densities 
the perturbative domain does not overlap with the densities one expects in a neutron star 
\cite{Kurkela:2014vha}. 
Despite the progress that has been made \cite{Kurkela:2009gj} in bringing perturbative calculations of
the properties of hot and dense QCD closer to the hadronic domain, the QCD phase transition from confined quarks in dense hadronic matter to deconfined quarks in a quark-gluon plasma
is characterized by features which are not accessible at any finite order of perturbative QCD\index{perturbative QCD}.
A standard example for this statement is the description of chiral symmetry breaking
which in the light quark sector is a distinctively non-perturbative feature.

The most common approach to avoid these difficulties is to apply effective quark matter models 
which are not derived from QCD but 
account for
certain characteristic features, most notably again the dynamical breaking of chiral symmetry.
Prominent examples for effective models are the thermodynamic bag model as the high density limit of the MIT bag model,
and effective relativistic mean field models, typically of the Nambu--Jona-Lasinio type
\cite{Buballa:2003qv}.
A next approach which is well developed for the study of hadrons is the non-perturbative Dyson-Schwinger formalism
which starts from the QCD action and derives gap equations to determine QCD's n-point Green-functions in a medium.
As these typically couple to higher order Green-functions, truncation schemes are introduced
which, if chosen wisely, preserve key features of QCD \cite{Cloet:2013jya}.

\subsection{Dynamical chiral symmetry breaking}

As described for hyperon matter, quark matter would be too soft to account for massive neutron
stars if repulsive interactions would not be taken into account. 
However, vector repulsion arises as naturally as the breaking of chiral symmetry in relativistic effective models and in the Dyson-Schwinger approach as vector and scalar part of the dressing (self-energy) of the fermion propagator.

The dynamical mass generation mechanism is common to all modern EoS models for quark matter in neutron stars. 
It can be obtained already with the simplest dynamical models that contain a 4-fermion interaction with sufficiently strong coupling in the scalar channel,  such as the celebrated Nambu--Jona-Lasinio (NJL) model\index{Nambu--Jona-Lasinio (NJL) model} of low-energy QCD that became a "workhorse" for quark matter phenomenology.
There are a few excellent reviews on the NJL model and its application to finite temperature and density, e.g., Refs. \cite{Klevansky:1992qe,Hatsuda:1994pi,Buballa:2003qv,Fukushima:2010bq,Fukushima:2013rx} to which we refer the reader for details. 

For definiteness, we like to show the typical form of an NJL model Lagrangian for applications to compact star physics as it is discussed in \cite{Baym:2017whm,Kojo:2014rca}. 
   The Lagrangian of the three-flavor NJL model is
\begin{equation}
\mathcal{L} = \overline{q} ( \gamma^\mu p_\mu  - \hat{m}_q +\mu_q \gamma^0) q + \mathcal{L}^{(4)} + \mathcal{L}^{(6)}   ,   
\label{eq:NJL_Lagrangian}
\end{equation}
where $q$ is the quark field operator with color, flavor, and Dirac indices, $\bar q = q^\dagger \gamma^0$, $\hat{m}_q$ the quark current mass matrix, $\mu_q$ the (flavor dependent) quark chemical potential and $\mathcal{L}^{(4)} = \mathcal{L}^{(4)}_\sigma +\mathcal{L}^{(4)}_d + \mathcal{L}^{(4)}_V$ and $\mathcal{L}^{(6)} = \mathcal{L}^{(6)}_\sigma + \mathcal{L}^{(6)}_{\sigma d}$ are four and six-quark interaction terms, chosen to reflect the symmetries of QCD.   

The first of the four-quark interactions, a contact interaction with coupling constant $G>0$,
\begin{equation}
\mathcal{L}^{(4)}_\sigma = G \sum^8_{j=0} \left[ (\overline{q} \tau_j q)^2 + (\overline{q} i \gamma_5 \tau_j q)^2 \right] = 8 G \mbox{tr} (\phi^\dagger \phi)   ,  
\label{eq:L4_sigma}
\end{equation}
describes spontaneous chiral symmetry breaking, where $\tau_j$ ($j= 0,\ldots,8$) are the generators of the flavor-U(3) symmetries, and in Eq.~(\ref{eq:L4_sigma}), 
\begin{equation}
\phi_{ij} = (\overline{q}_R)^j_a (q_L)^i_a
\end{equation}
is the chiral operator with flavor indices $i,j$ (with summation over the color index $a$); the right and left quark chirality components are defined by  $q_{R,L} = \frac12(1\pm \gamma_5)q$.

The second of the four-quark terms describes the scattering of a pair of quarks in the s-wave, spin-singlet, flavor- and color-antitriplet channel; this interaction leads to BCS pairing of quarks:
\begin{eqnarray}
\mathcal{L}^{(4)}_d & = & H \sum_{A,A^\prime = 2,5,7} \big[ \left(\overline{q} i \gamma_5 \tau_A \lambda_{A^\prime} C \overline{q}^T \right) \left(q^T C i \gamma_5 \tau_A \lambda_{A^\prime} q \right)   \nonumber   \\
&& \hspace{23mm} + \left(\overline{q} \tau_A \lambda_{A^\prime} C \overline{q}^T \right) \left(q^T C \tau_A \lambda_{A^\prime} q \right) \big]   ,   \nonumber   \\
& = & 2 H {\rm tr} (d^\dagger_L d_L + d^\dagger_R d_R),
\label{eq:L4_d}
\end{eqnarray}
with $H>0$.  Here $\tau_A$ and $\lambda_{A^\prime}$ ($A,A'=2,5,7$) are the antisymmetric generators of U(3) flavor and SU(3) color, respectively, and 
\begin{equation}
(d_{L,R})_{ai} = \epsilon_{abc} \epsilon_{ijk} (q_{L,R})^j_b {\cal C} (q_{L,R})^k_c 
\label{diquark_field}
\end{equation}
are diquark operators of left- and right-handed chirality,  with ${\cal C}=i\gamma^0\gamma^2$ the charge conjugation operator. 
The diquark pairing\index{diquark pairing} interaction leads as well to an attractive correlation between two quarks inside confined hadrons and, in constituent quark models, plays a role in
the observed mass splittings of hadrons \cite{DeRujula:1975qlm,Anselmino:1992vg,Jaffe:2004ph}. 
This interaction, in weak coupling, arises from single gluon exchange;  however at the densities of interest in neutron stars,  the non-linearities of QCD prevent direct calculation of this interaction, and so one must treat it phenomenologically.

In addition
\begin{eqnarray}
\mathcal{L}_V^{(4)}  =  -g_V (\overline{q} \gamma^\mu q)^2   ,   
\label{vec}
\end{eqnarray}
with $g_V > 0$, is the Lagrangian for the phenomenological vector interaction, 
which produces universal repulsion between quarks \cite{Kunihiro:1991qu}. 

The six-quark interactions represent the effects of the instanton-induced QCD axial anomaly, which breaks the $U(1)_A$ axial symmetry of the QCD Lagrangian. 
The resulting Kobayashi-Maskawa-'t Hooft (KMT) interaction\index{Kobayashi-Maskawa-'t Hooft (KMT) interaction} leads to an effective coupling between the chiral and diquark condensates of the form~\cite{Kobayashi,tHooft:1986ooh}:
\begin{eqnarray}
\label{eq:L6_sigma}   
\mathcal{L}^{(6)}_\sigma &=& -8 K (\det \phi + \mbox{h.c.}),\\
\mathcal{L}^{(6)}_{\sigma d} &=& K^\prime ( {\rm tr} [(d^\dagger_R d_L) \phi] + \mbox{h.c.})   ,   
\label{eq:L6_sigma_d}
\end{eqnarray}
where $K$ and $K^\prime$ are positive constants. Provided that $K^\prime \simeq K$ (which one expects on the basis of the Fierz transformation connecting the corresponding interaction vertices) the six-quark interactions encourage the coexistence of the chiral and diquark condensates.  

Before discussing the hadron-to-quark matter phase transition under neutron star constraints 
in Sect.~\ref{sec:HQPT} and the phenomenology of hybrid compact stars with quark matter interior 
in Sect.~\ref{sec:HSP} we briefly discuss some of the challenges in describing deconfined quark matter
in the nonperturbative domain of the QCD phase diagram that borders the hadronic matter phases.

\subsection{Color superconductivity\index{color superconductivity}}

Quark matter at finite densities and zero or small temperature can exhibit an extremely rich
phase structure due to different pairing mechanisms which arise from the coupling of
color, flavor and spin degrees of freedom and result in a variety of different possible condensates
\cite{Alford:2001dt,Alford:2007xm}.
The importance of condensates is illustrated by the color-flavor locked phase which can appear 
in three flavor (up, down, strange) matter and is shown to be the asymptotic ground state of quark matter
at low temperature \cite{Schafer:1999jg}.
This zoo of color superconducting phases
bears a rich potential for the phenomenology, in particular for the cooling of compact stars with color superconducting quark matter cores. 
While for applications in hybrid star cooling it is the smallest diquark pairing gaps that matter most
\cite{Grigorian:2004jq}, for direct effects on the EoS the largest gaps are most important.
Those are usually found in the scalar diquark channels where they can be of the order of the dynamically generated quark masses and thus result in effects on the EoS as important as chiral symmetry breaking.
For the discussion of the QCD phase diagram it is therefore crucial to solve the coupled gap equations for masses and diquark gaps simultaneously. 
For the three-flavor case this has been accomplished first in \cite{Ruester:2005jc,Blaschke:2005uj,Abuki:2005ms}. 
For usual NJL model\index{Nambu--Jona-Lasinio (NJL) model} parametrisations critical temperatures for onset of color superconductivity around 50 MeV arise. 
When quarks get coupled to the 
Polyakov-loop\index{Polyakov-loop} and thus their excitation is strongly suppressed in the confined phase and still moderately suppressed in the deconfined one, the critical temperature for the two-flavor superconducting (2SC) phase\index{2SC phase} can become as high as 150 MeV and thus be of the order of the chiral restoration temperature and temperatures for chemical freeze-out in heavy-ion collisions, see
\cite{Blaschke:2010ka,Abuki:2008nm}.
It is worth to highlight that the restoration of chiral symmetry is flavor dependent. 
This results in a sequential appearance of quark flavors at different densities 
\cite{Blaschke:2008br,Alford:2017qgh}, with the almost simultaneous occurrence of the corresponding color superconducting phase - 2SC for two-flavor quark matter and color-flavor-locking (CFL)\index{CFL phase} for three-flavor quark matter. With a neutralizing surrounding of positively charged hadrons even a  single flavor phase phase is possible due to an effect analogous to the neutron drip line - the down-quark dripline occurs and these down quarks can obey a single flavor pairing pattern resulting in the color-spin-locking (d-CSL) phase\index{d-CSL phase} \cite{Blaschke:2008br}.   

In the case of a strongly asymmetric occupation of the Fermi levels of different quark flavors, as in the case of $\beta-$equilibrated compact star matter, an energetically favorable ground state is found when the diquark Cooper pairs have finite momentum. One speaks of crystalline phases\index{crystalline phases} of QCD or, in analogy to such a situation in the electronic superconductors in a magnetic field, of the Larkin-Ovchinnikov-Fulde-Ferrell (LOFF) state\index{Larkin-Ovchinnikov-Fulde-Ferrell (LOFF) state} of matter. An example of (rotating) hybrid star sequence with crystalline color superconducting quark matter cores fulfilling also the modern maximum mass constraint has been presented in Ref.~\cite{Ippolito:2007hn}. For a review on crystalline phases of QCD, see \cite{Anglani:2013gfu}.

\begin{figure}
	\centering
		\includegraphics[width=0.8\textwidth]{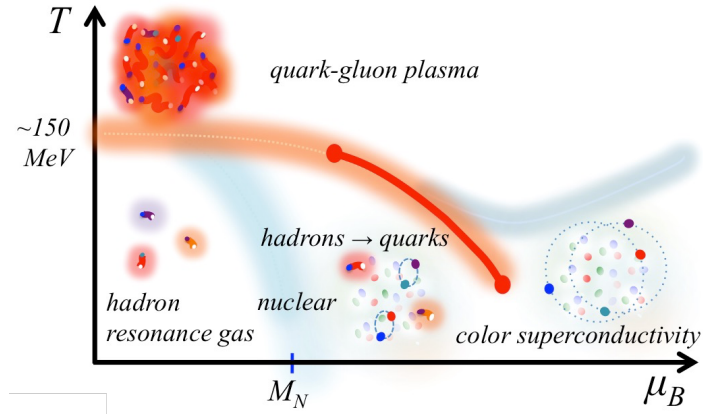}
	\caption{Schematic phase diagram\index{QCD phase diagram} of dense nuclear matter with two critical endpoints of first order 
		phase transitions, from Ref.~\cite{Kojo:2014rca}.
		The crossover at low temperatures is due to the BEC-BCS crossover \cite{Hatsuda:2006ps,Abuki:2010jq}. 
	}
	\label{fig:PhD-Baym}       
\end{figure}

Usually the quark mass gap and the diquark pairing gap repel each other, but for strong coupling it is possible that a coexistence of chiral symmetry breaking and color superconductivity occurs which changes the order of the chiral restoration transition from first order to crossover\index{crossover transition}.
A particularly interesting scenario is the crossover transition of this kind which results from the coupling of chiral and diquark gaps via the six-quark interaction term (\ref{eq:L6_sigma_d}) that stems from the Fierz transformed KMT determinant interaction\index{Kobayashi-Maskawa-'t Hooft (KMT) interaction} \cite{Hatsuda:2006ps,Abuki:2010jq}, 
see Fig.~\ref{fig:PhD-Baym} for an illustration of this case which is in accordance with the idea of quark-hadron continuity\index{quark-hadron continuity} \cite{Schafer:1998ef} or more general parton-hadron duality \cite{Wetterich:1999vd}.
  
The question arises whether such a picture of a crossover transition at zero temperature in compact stars would exclude characteristic signatures of a strong first order phase transition in the mass-radius diagram such as mass twin stars and a third family sequence. We come back to this question in Sect. \ref{sec:HSP} below.  

\subsection{Stiffness} 

The stiffness of quark matter in NJL type models is determined by the coupling of quarks to a vector mean-field which has a self-interaction resulting from the 4-fermion coupling (\ref{vec}) in the model Lagrangian, analogous to the linear Walecka model for nucleons.
The inclusion of this coupling was essential for obtaining stable branches of hybrid stars with masses 
reaching up to and above $2~M_\odot$ when  a NJL type model was used for the quark matter EoS, see \cite{Klahn:2006iw} for a first paper fulfilling this constraint. 
While increasing the vector coupling $g_V$ raises the maximum mass of the hybrid star sequence, increasing the scalar diquark coupling $H$ decreases it, together with a decrease in the onset mass for the deconfinement transition.  
A systematic study of the parameter space has been performed in \cite{Klahn:2013kga} which showed that stable hybrid stars which reach up to $2~M_\odot$ could be obtained with a NJL model quark matter core and a hadronic shell described by the DBHF\index{Dirac-Brueckner-Hartree-Fock (DBHF)} EoS with the 
Bonn-A interaction only when both, the vector coupling and the diquark coupling were nonvanishing and sufficiently strong.
  
Recently, an eight-quark interaction\index{eight-quark interaction} term in the vector channel, 
\begin{equation}
\mathcal{L}_V^{(8)}=-G \eta_4 (\bar{q} \gamma_\mu q)^4,
\label{L8V}
\end{equation} 
has been introduced in \cite{Benic:2014iaa} in order to describe a stiffening of quark matter at high densities\index{stiffness of the EoS} which improves the stability of compact hybrid star configurations against gravitational collapse  \cite{Benic:2014jia}. 
Such an interaction is indispensable when one wants to describe within a microscopically motivated EoS model the occurrence of a third family of hybrid stars fulfilling the $2~M_\odot$ mass constraint.
Such a higher order quark interaction can be motivated by the existence of vector boson couplings of fourth order. Such boson self-coupling terms occur in the nonlinear Walecka model, but they are also important for explaining a sufficient repulsion in the high-energy nucleon-nucleon scattering by multi-pomeron exchange\index{multi-pomeron exchange}
\cite{Rijken:2016uon}. 

\subsection{Confinement\index{confinement}}

The NJL model finds wide application as a microscopic model for quark matter in compact stars since in this case at $T=0$ the dynamical chiral symmetry breaking mechanism also mimics confinement (see, e.g., \cite{Klahn:2013kga,Hell:2014xva} and references therein). 
This is because at the mean field level the value of the constituent quark mass is independent of the chemical potential $\mu$ (as well as the pressure $P$) so that the quark density $n=\partial P/\partial \mu$ vanishes as long as $\mu$ stays below its critical value  $\mu_c$.
For $\mu>\mu_c$ the quark mass jumps down and the pressure evolves with $\mu$ so that there is a finite quark density. 
However, the NJL model\index{Nambu--Jona-Lasinio (NJL) model} has no true confinement and this becomes apparent at finite temperatures, where the partial pressure and density of the quarks becomes of the same order as the hadronic one already for temperatures below the pseudocritical temperature $T_c=154\pm 9$ MeV known from lattice QCD simulations.
This problem is severe when applications for finite temperature systems are to be considered like  protoneutron stars in supernova simulations or hot, hypermassive neutron stars in neutron star merger events.     
Unfortunately, the minimal coupling of the Dirac quarks to a homogeneous gluon background field within the Polyakov-loop improved NJL (PNJL) model would not be an appropriate approach to account for confining effects in $T=0$ quark matter since we have no possibility to calibrate a chemical potential dependence of the Polyakov-loop\index{Polyakov-loop} potential\footnote{An ansatz for the Polyakov-loop potential which is applicable also at $T=0$ has been suggested by Dexheimer \& Schramm \cite{Dexheimer:2009va}. For an application to hybrid stars, see \cite{Blaschke:2010vj}} with lattice QCD simulations, unlike the case at vanishing baryon density and finite temperatures where the PNJL model is an acceptable model for low-energy QCD matter \cite{Ratti:2005jh,Roessner:2006xn}.

A standard minimal way do account for quark confinement in a  thermodynamic description is the thermodynamic bag model\index{thermodynamic bag model} which consists in adding to the pressure of a relativistic Fermi gas of quarks (plus Bose gas of gluons, eventually improved by perturbative corrections of $\mathcal{O}(\alpha_S)$) a negative pressure contribution $P_{\rm bag}=-B$, with $B>0$ being the bag constant\index{bag constant}. 
This makes sure that at low temperatures and chemical potentials always the hadronic matter dominates the thermodynamics because it has a nonnegative pressure. 
The nonperturbative bag pressure can be understood to originate from quark and gluon condensates,
$B=B_q + B_g$. 
The NJL model allows to calculate the medium-dependent value of a chiral condensate 
$B_q \propto \langle \bar{q} q\rangle$ \cite{Buballa:2003qv} and fits for the $\mu-$dependent quark bag function $B_q(\mu)$ from a nonlocal color superconducting have been provided for hybrid neutron star calculations \cite{Grigorian:2003vi}. Qualitatively, the density dependent bag function starts at a finite value and gets lowered with increasing density. Such functions have been modelled on heuristic grounds also in Refs.~\cite{Burgio:2001mk,Burgio:2002sn} and may be regarded as a density functional.
Since typical NJL-type models have no gluon dynamics, an understanding of the gluonic contribution to the bag function and its medium dependence is lacking. Again qualitatively, one can expect that the gluon condensate should melt at increasing temperature and density such that a positive pressure contribution should result.    
Such bag-like contributions to the pressure of NJL-model quark matter have been introduced, e.g. in \cite{Pagliara:2007ph,Bonanno:2011ch,Blaschke:2010vj}, where due to this addition a stable hybrid star with a CFL quark matter\index{CFL phase} core could be obtained. In NJL quark matter models without such a contribution the hybrid stars turn gravitationally unstable as soon as the strange quark flavor appears \cite{Klahn:2006iw,Alford:2006vz,Klahn:2013kga}.
Another argument to add a (medium dependent) bag constant to an NJL-type model is to enforce coincidence of the chiral restoration transition in the QCD phase diagram\index{QCD phase diagram} with the hadron-to-quark matter transition, see \cite{Klahn:2015mfa}.
In these models an appropriately chosen vector interaction (\ref{vec}) has to be added in order to achieve a  sufficient stiffening of the resulting hybrid star EoS in order to fulfill the $2~M_\odot$ constraint.

However, since bag type models, even if improved by coupling to vector and scalar mean fields, have no dynamical confinement 
we should therefore advance to models that make use of a confining interquark potential, like the  
field correlator method (FCM)\index{field correlator method (FCM)}, see \cite{Logoteta:2013aca} and references therein. 
This model constructs from color electric and magnetic correlators a mean vector field that contributes 
an energy shift to the quark and gluon distribution functions and strongly suppresses them at low densities. This model reproduces rather well a constant speed of sound (CSS) model\index{constant speed of sound (CSS) model} for quark matter with, however, a low squared speed of sound $c_s^2\sim 1/3$ \cite{Alford:2015dpa}. 
This model has the strength that at finite temperatures its input parameters (gluon condensate $G_2$ and vector field $V_I$) can be fixed with lattice QCD. 
However, when these values are used for the construction of hybrid stars there is a problem to fulfill the $2~M_\odot$ constraint \cite{Pereira:2011yq}.  Moreover, no absolutely stable strange quark matter is obtained \cite{Pereira:2012tr}.

Another possibility to include confining effects into a field-theoretic description of quark matter is the color dielectric model (CDM)\index{color dielectric model (CDM)} \cite{Alberico:2001zz}. 
As opposed to the FCM the color dielectric field $\chi$ is a scalar and determines the quark masses.
At low densities and thus low values of $\chi$ the quark masses diverge and this suppresses the excitation of these degrees of freedom. 
Interestingly in the CDM, like in the bag model, absolutely stable strange quark matter is possible.
Maieron et al.~\cite{Maieron:2004af} showed that for the CDM and a bag model with density-dependent B the maximum masses of hybrid stars are never larger than $1.6~M_\odot$.

Very recently, a relativistic density functional approach to quark matter has ben developed \cite{Kaltenborn:2017hus} which implements a confining mechanism in the spirit of the CDM in its density functional which in this case is motivated by the string-flip model\index{string-flip model (SFM)} of quark matter \cite{Ropke:1986qs} and is a generalization of \cite{Li:2015ida} and an earlier density functional approach developed for heavy-ion collision applications \cite{Khvorostukin:2006aw}. We will refer to this approach as SFM below. 
Its density functional subsumes all aspects required for the definition of a class of hybrid star EoS which may fulfill all known constraints from the phenomenology of compact stars.

\subsection{Strange Quark Matter in Neutron Star Cores}

We first discuss the situation where normal nuclear or hyperon matter deconfines at a critical density and forms a quark-gluon plasma.
This problem is still far from being solved consistently in a way where hadrons would be taken into account as actually confined quarks which then deconfine dynamically.
The standard way to circumvent a detailed description of deconfinement is to choose a two phase approach
where nuclear and quark matter are modeled independently and the transition is constructed thermodynamically consistently, viz. in terms of a Maxwell (or similar) construction.
In case of the Maxwell construction\index{Maxwell construction} one simply determines at which baryon-chemical potential the pressure of the nuclear and quark phase are equal and thus defines the transition point.
By construction, this implies a first order phase transition\index{first order phase transition} which 'switches' from a given nuclear equation of state to
a softer quark matter equation of state. Therefore, this procedure requires a nuclear EoS which is stiff enough to support at least a two solar mass neutron star, and a quark matter EoS which is softer but stiff enough to do the same.
How exactly this happens can vary. 
The quark matter EoS can mimic the nuclear EoS, be generally softer or at some
density can turn even stiffer than the underlying nuclear EoS
so that a second crossing of the pressure curves occurs ("reconfinement problem"\index{reconfinement problem} \cite{Zdunik:2012dj}). 
The latter scenario is justified if one assumes that
at densities far enough beyond the transition a comparison of both phases is meaningless as the nuclear EoS does not describe any physical reality anymore.
In any of these scenarios, repulsion is a crucial feature to account for the existence of massive neutron stars and, as stated earlier, a natural property of relativistic models.
For the onset density of hyperon matter\index{hypernuclear matter} we discussed how it is pushed to increasingly high densities with increasing stiffness\index{stiffness of the EoS} or repulsion.
The same would be true for quark matter if condensates are not taken into account.
However, condensates couple colors and flavors and can lower
the transition density. 
Therefore, quark models can account for transition densities far below the hyperon threshold.

For the appearance of strange matter in neutron stars, a couple of scenarios emerge which we divide into two groups.
First, nuclear matter can deconfine directly into (three flavor) strange matter as one would find it for the thermodynamical bag model, as described earlier.
For a dynamical treatment of this transition as a nucleation process see, e.g., Refs.~\cite{Bombaci:2016xuj,Bhattacharyya:2017mdh} and references therein.
Second, a sequential transition\index{sequential transition} from nuclear to two-flavor followed by a transition to three-flavor matter takes place.
Similar to this scenario, there are two more cases which would not result in a strange matter core but are not less realistic.
The sequential transition results in neutron star configurations which are stable at all densities
below the strange quark threshold but unstable beyond due to the softening of the EoS with this new degree of freedom.
This situation has been described for an NJL model with diquark couplings 
\cite{Klahn:2006iw}.
However, choosing a finite constant as offset to the quark pressure
can alter this result \cite{Bonanno:2011ch} and render neutron stars with strange matter core stable 
(the thermodynamic properties of matter are described in terms of pressure derivatives, hence a constant offset keeps them intact).
Another way to stabilize strange core configurations is a strongly density dependent stiffening
of the strange matter EoS following the transition
\cite{Benic:2014jia,Kaltenborn:2017hus}.
This can result in situations where stable two flavor and three flavor quark core neutron star configurations are separated by a population gap at intermediate central neutron star densities.
In extreme scenarios, this can generate separated mass twin\index{mass twins} configurations, viz.
two neutron star families with similar masses but very different radii 
\cite{Blaschke:2013ana,Alvarez-Castillo:2017qki}.
If members of both families could be observed, this would be a strong indicator
for a first order QCD-like phase transition where (the smaller) compact stars can carry a core made of strange matter \cite{Bastian:2017fzo}. 
Note, however, that according to the two-families scenario\index{two families scenario} of Ref.~\cite{Drago:2013fsa} the smaller stars would be the hadronic ones (with a maximum mass of 1.5-1.6 M$_\odot$) while the very massive ones would be strange stars with a larger radius. 
They can coexist in this scenario because the hadronic stars are metastable with respect to the nucleation of strange quark matter.

Another indicator for a first order phase transition\index{first order phase transition} in dense matter would be the observation of a delayed second neutrino signal after a supernova. 
This has been suggested based on simulations which applied a bag model \cite{Sagert:2008ka,Fischer:2010wp}.
Although such a measurement would be very exciting it is not clear how it would address the question whether the transition involved strange matter, 
or quark matter at all.

\subsection{Absolutely Stable Strange Matter?\index{strange quark matter}}

The hypothesis that strange matter could be absolutely stable, bases 
on the observation that the appearance of strange quarks lowers the energy per baryon.
As two flavor quark matter is evidently less stable than Fe (otherwise Fe would decay into it's quark components and so would we)
this leaves a window where two-flavor matter is less and strange matter more stable than iron \cite{Bodmer:1971we,Witten:1984rs,Farhi:1984qu}.
This hypothesis has been supported by certain parametrisations of the thermodynamic bag model\index{thermodynamic bag model}.
Choosing the proper bag constant\index{bag constant} one can indeed find exactly the proposed situation.
If this scenario is reality it would have a number of interesting consequences.
Therefore, the search for stable strange matter inspired a multitude of experiments and has born many new ideas.
Strange matter could form strange nuggets of extreme density with rather small atomic numbers and hence
extremely low cross sections. It could form objects very similar to a neutron star, almost entirely made of strange matter, see~\cite{Bombaci:2001uk,Weber:2004kj} for reviews of this and other scenarios involving strangeness in compact stars.
A seed of strange matter in a neutron star could destabilize the surrounding matter and thus trigger a conversion of the neutron star interior into strange matter. 
Recently, it has been proposed that muonic bundles that have been observed at ALICE (CERN) are produced by strangelets \cite{Kankiewicz:2017wbo}.
Currently, other ways to detect strangelets are actively investigated \cite{VanDevender:2017avk}.
The idea of absolutely stable strange matter is certainly appealing.

\begin{figure}
	\centering
	\includegraphics[width=0.8\textwidth]{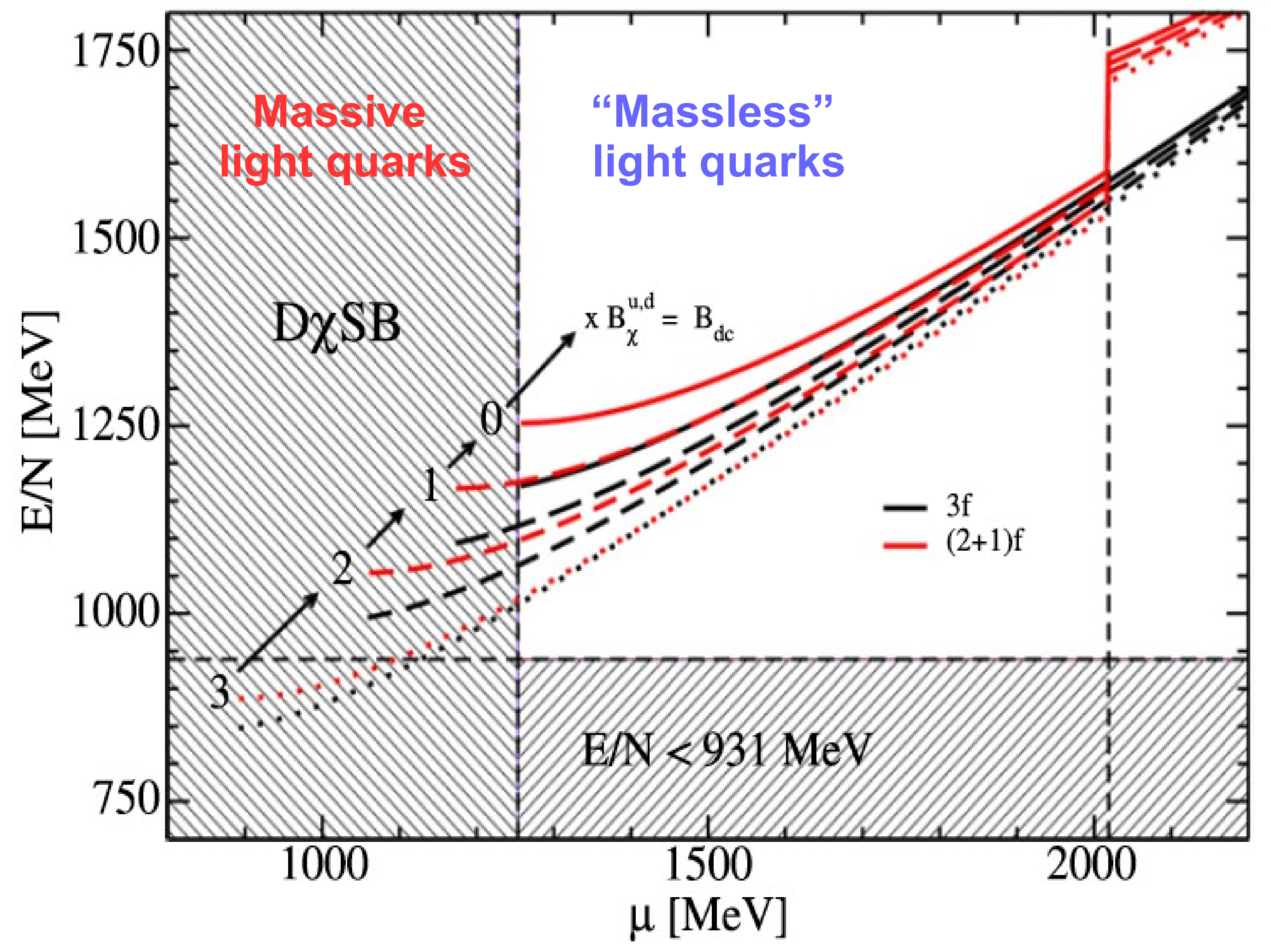}
	\caption{Energy per baryon vs. baryon chemical potential for the vBag quark matter model
		\cite{Klahn:2015mfa}. 
		Absolutely stable strange quark matter\index{strange quark matter} could be obtained if dynamical chiral symmetry breaking (D$\chi$SB) is neglected. However: The vertical band  marks the approximate region where chiral symmetry is broken. For the shown curves D$\chi$SB is ignored for all (black) or all but the s-quark (red).
		At densities below the s-quark threshold this allows to effectively compare two(red)- and three-flavor(black) quark matter for different effective bag constants. In none of the cases strange matter is more stable than iron ($E/N < 931$ MeV) in a density domain where chiral symmetry is restored. See also Ref.~\cite{Klaehn:2017mux}.}
	\label{fig-1}       
\end{figure}

Theoretically, as mentioned before, the hypothesis has been based on the thermodynamic bag model.
It should be noticed, that this model lacks a key feature of QCD, namely chiral symmetry breaking, 
viz. quarks in the thermodynamic bag model are assumed to have bare masses and therefore
extremely small threshold densities (the critical chemical potential scales with the effective mass).
NJL-type models\index{Nambu--Jona-Lasinio (NJL) model}, which do generate dressed quark masses, do not confirm that strange
matter is absolutely stable. 
The reason for this is easily found: In the (low) density domain where the bag model predicts
absolutely stable strange matter\index{strange quark matter}, NJL type models find chiral symmetry to be broken, hence significantly larger quark masses and consequently a higher energy per particle, too high to render strange matter absolutely stable, see Fig.~\ref{fig-1}. 
An appealing and sometimes confusing feature of the thermodynamic bag model is that it originates from
the MIT bag model\index{MIT bag model} which has been developed to describe hadron properties. 
This evidently seems possible even though the model assumes bare quark masses
and makes it easier to believe that massless quarks could form stable strangelets.
This confusion can be cleared if one realizes that the MIT bag model does not only
assume bare quark masses but a bag constant\index{bag constant} which is introduced as synonym for 'all we don't know about confinement and further interactions'. 
The fathers of the MIT bag model stated this explicitly.
In \cite{Klahn:2015mfa} we illustrated how to translate a model with chiral symmetry breaking
into a bag type model and where this model would break, which is when chiral symmetry is broken.
For an earlier investigation of this kind, see \cite{Buballa:2003qv} for the NJL model case and  
\cite{Gocke:2001ri} for a nonlocal chiral quark model. 
Effects of color superconductivity in this context have been studied and parametrized in  \cite{Grigorian:2003vi}.
%

The bag constant mostly originates from chiral symmetry breaking. It is subtracted because the difference between the vacuum pressure of massive quarks and effectively massless quarks is negative. The absolute value is reduced by confinement. Thinking of the absolute value of the bag constant as the energy of a hadron, this makes sense as 
confinement reduces the energy of the system of chirally broken quarks by the binding energy.
The MIT bag model is not consistent in the sense that it ignores any relation
between quark mass and bag constant. In vacuum this is a reasonable approximation for two reasons:
First, the model describes hadrons and does not attempt to predict any dynamical property of individual constituent quarks.
Second, it can be fitted to observables and thus repairs the inherent shortcomings.
The thermodynamic bag model\index{thermodynamic bag model} addresses both effects, chiral symmetry breaking and confinement, in terms of one parameter - the bag constant. Statements regarding quark matter based on the thermodynamic bag model, in particular at low densities where the bag constant affects the total pressure significantly, should be considered with considerable caution as it is all but clear, that the model actually describes deconfined matter.

It should be noticed, that effective quark models with density dependent quark masses 
have been suggested which indeed would predict absolutely stable strange matter
\cite{Dondi:2016yjl,Bastian:2017}.
A distinct feature of these models is a steep {\it concave} decrease of the strange quark mass at
comparably low density opposing to the typically {\it convex} behavior. 
This reduces the effective quark mass drastically already at low densities 
which makes it very similar to to the thermodynamic bag model,
evidently with similar results regarding the stability of strange matter.
It would be interesting to see, how a microscopic approach which generates
this kind of density behavior would perform describing hadron properties.

\section{Hadron-to-quark matter phase transition}
\label{sec:HQPT}


\subsection{Maxwell construction\index{Maxwell construction} \& beyond}

When the hadronic and quark matter phases are described with EoS given by relations between the pressure and chemical potential (for $T=0$, which is relevant for the NS modeling)
$P_{H}(\mu)$ and $P_{Q}(\mu)$ 
correspondingly, one can find the critical value of the baryochemical potential $\mu_c$ from the condition of equal pressures, 
\begin{equation}
P_{Q}\left(\mu_{c}\right)=P_{H}\left(\mu_{c}\right) = P_c,
\end{equation}
for which the phases are in mechanical equilibrium with each other.  
The value $P_c$ defines the Maxwell construction. 
Quantities characterizing the quark-, hadron-, and mixed phases are denoted by the subscripts $Q$, $H$, and $M$, respectively.

Assuming the surface tension to be smaller than the critical value $\sigma_c$,
a mixed phase could have an influence on the compact stars structure. 
The adequate description of the physics of pasta phases\index{pasta phases} is a complicated problem which requires to take into account sizes and shapes of structures as well as transitions between them.
It has been dealt with in the literature within different methods and approximations
\cite{Voskresensky:2002hu,Yasutake:2014oxa,Maruyama:2007ss,Watanabe:2003xu,Horowitz:2005zb,Horowitz:2014xca,newton2009}.

\begin{figure}[!tb]
	\includegraphics[scale=0.4]{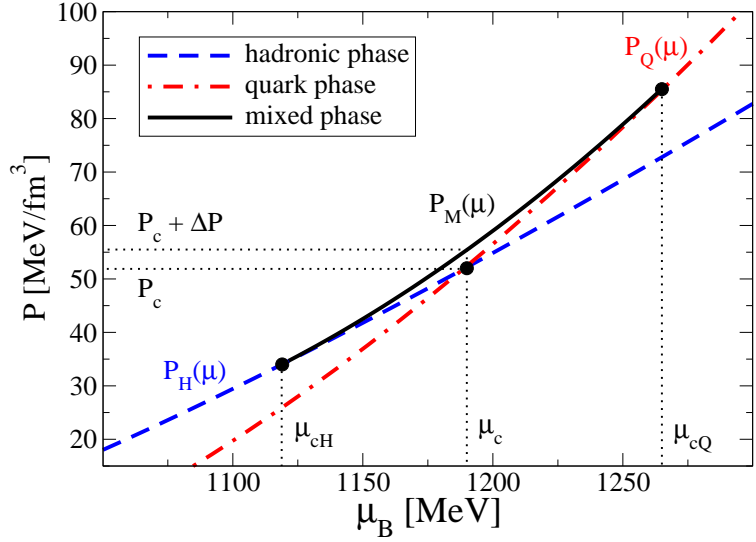}
	\caption{Maxwell construction of a first-order phase transition between a hadronic EoS $P_H(\mu)$ and a quark matter EoS $P_Q(\mu)$ occurring at the critical pressure $P_c=P(\mu_c)$. 
    The modification by a parabolic function $P_M(\mu)$ mimicks pasta phases.
		Adapted from Ref.~\cite{Ayriyan:2017nby}.}
	\label{fig:pasta}
\end{figure}

Here, instead of the full solution we would like to use a simple modification  of the Maxwell construction which mimics the result of pasta matter studies \cite{Ayriyan:2017tvl}.  
This means that the pressure as a function of the baryon density in the mixed phase is not constant 
but rather a monotonously rising function. 
%
We are assuming that close to the phase transition point (that would be obtained using the Maxwell construction) the EoS of both phases are changing due to finite size and Coulomb effects (see Fig.~\ref{fig:pasta}), so that the effective mixed phase EoS $P_{M}(\mu)$ could be described in the parabolic form 
\begin{equation}
P_{M}(\mu)=a (\mu-\mu_{c})^2+ b (\mu-\mu_{c})+P_c+\Delta P.
\label{p_mu}
\end{equation}
Here we have introduced the pressure shift $\Delta P$ at $\mu_{c}$ as a free parameter of the model
which determines the mixed phase pressure at this point
\begin{equation}
P_{M}\left(\mu_{c}\right)=P_c+\Delta P = P_M.
\end{equation}
{
	The parameter $\Delta P$ shall be related to the largely unknown surface tension between the hadronic and quark matter phases of the strongly interacting system at the phase transition. Infinite surface tension corresponds to the Maxwell construction and thus to $\Delta P=0$, while a vanishing surface tension is the case of a Gibbs construction\index{Gibbs construction} under global charge conservation (also called Glendenning construction \cite{Glendenning:1992vb}) that for known examples looks similar to the results of our mixed phase construction for the largest values considered here, $\Delta_P\sim 0.07 \dots 0.10$.
	A more quantitative relation between $\Delta P$ and the surface tension would require a fit of the mixed phase parameter to a pasta phase calculation for given surface tensions. Such a calculation can be done along the lines of Ref.~\cite{Yasutake:2014oxa}. 
}

According to the mixed phase construction\index{mixed phase construction} shown in Fig.~\ref{fig:pasta} we have two critical chemical potentials, 
$\mu_{cH}$ for transition from $H$-phase to $M$-phase and $\mu_{cQ}$ for the transition from the 
$M$-phase to $Q$-phase. 
Together with  the coefficients $a$ and $b$ from Eq.~\eqref{p_mu} this are four unknowns which 
shall be determined from the four equations for the continuity of the pressure 
\begin{eqnarray}
\label{PH}
P_{M}(\mu_{cH}) & = & P_{H}(\mu_{cH}) = P_H,\\
P_{M}(\mu_{cQ}) & = & P_{Q}(\mu_{cQ}) = P_Q,
\label{PQ}
\end{eqnarray}
and of the baryon number density $n(\mu)=dP(\mu)/d\mu$ 
\begin{eqnarray}
\label{nH}
n_{M}(\mu_{cH}) & = & n_{H}(\mu_{cH}),\\
n_{M}(\mu_{cQ}) & = & n_{Q}(\mu_{cQ}).
\label{nQ}
\end{eqnarray}

From the Eqs.~ (\ref{PH}) and (\ref{PQ}) for the pressure $a$ and $b$ are found 
and can be eliminated from the set of equations for the densities.
By solving the remaining Eqs.~ (\ref{nH}) and (\ref{nQ}) for the densities numerically one can find the values for the critical chemical potentials $\mu_{cH}$ and $\mu_{cQ}$.

\begin{figure}[!tb]
	\includegraphics[scale=0.35]{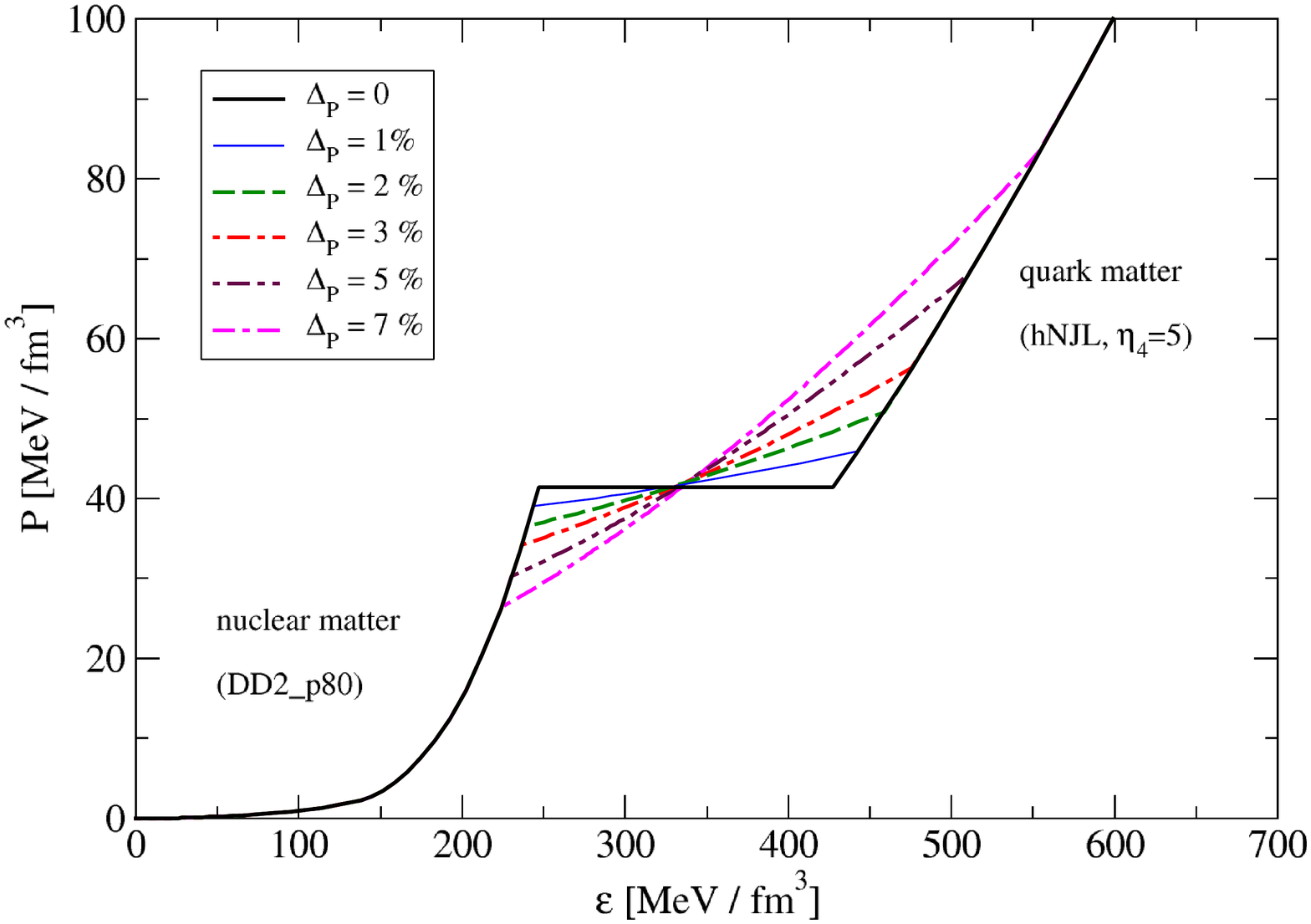}
	\includegraphics[scale=0.35]{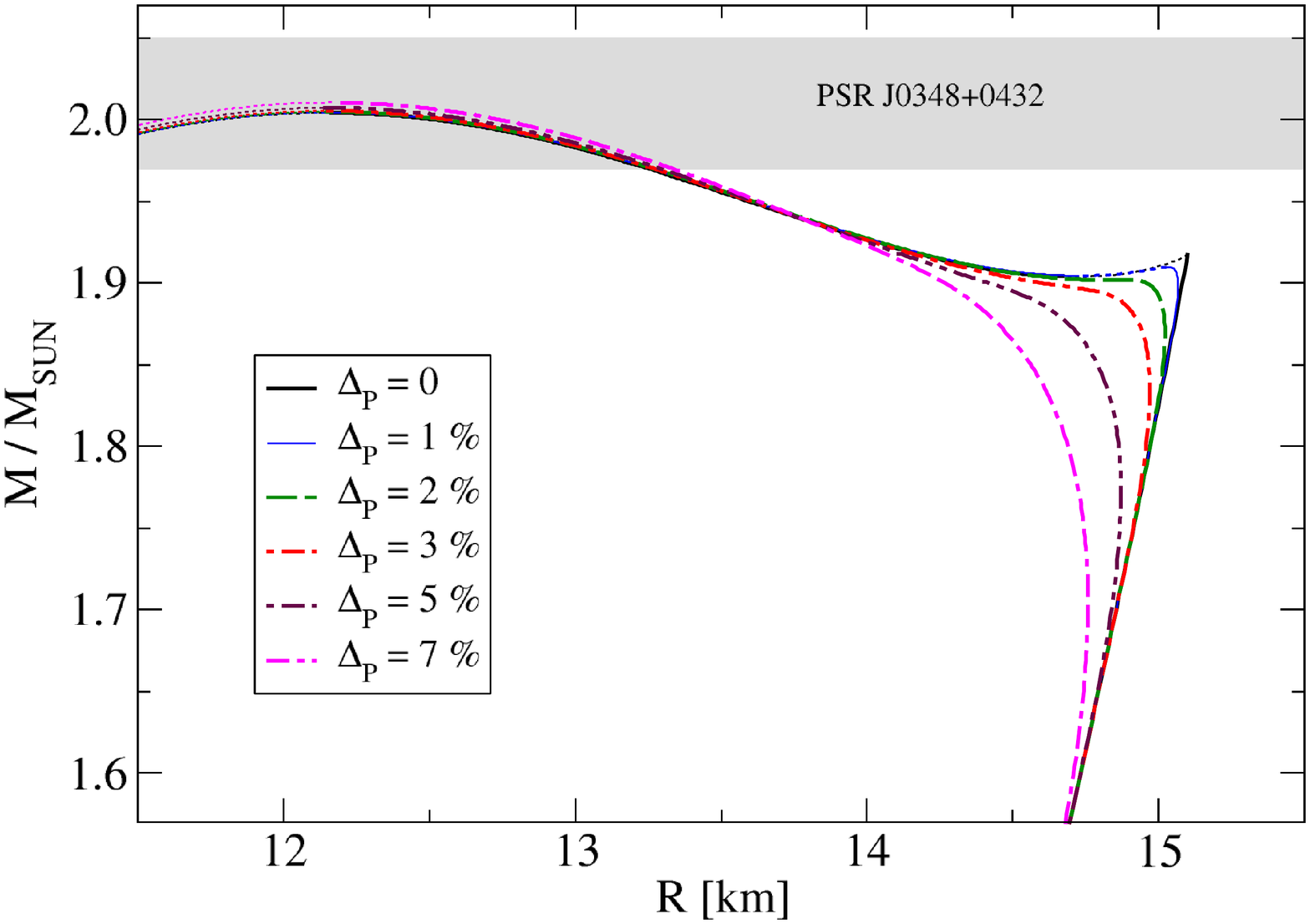}
	\caption{Two-phase model for the hadron-to-quark matter transition obtained by a Maxwell construction ($\Delta_P=\Delta P/P_c =0$) and its modifications ($\Delta_P=1,2,3,5,7 ~\%$) which mimic mixed phases with pasta structures. Upper panel: $P$ vs. $\varepsilon$ for the EoS from \cite{Benic:2014jia} where the nuclear matter phase is DD2 with excluded volume and the quark matter phase is a NJL model with 8-quark interactions. Lower panel: Solutions of the TOV equations for this EoS show that for $\Delta_P>1~\%$ the sequence of hybrid star solutions does not form a third family separated from that of the neutron star ones. 
		Adapted from Ref.~\cite{Ayriyan:2017nby}.}
	\label{fig:EoSpastaMR}
\end{figure}

In Fig~\ref{fig:EoSpastaMR} we show results of this mixed phase construction for the example of EoS investigated in Benic et al. \cite{Benic:2014jia} for different values of the parameter $\Delta_P$
(upper panel) and the corresponding solutions of the TOV equations (lower panel).
With this modification of the Maxwell construction one can examine the robustness of the third family solutions against pasta phase effects. Further details and other EoS combinations, see 
\cite{Ayriyan:2017nby,Alvarez-Castillo:2017xvu,Ayriyan:2017tvl}  

The procedure for "mimicking pasta phases" described in this subsection is general and can be 
superimposed to any hybrid EoS which was obtained using the Maxwell construction of a first-order
phase transition from a low-density to a high-density phase with their given EoS.
A certain limitation occurs in the case of sequential phase transitions (see, e.g.,  
Refs.~\cite{Blaschke:2008br}  and \cite{Alford:2017qgh}) when the broadening of the phase transition region as shown in Fig.~\ref{fig:pasta} would affect the adjacent  transition. 

\begin{figure}[!htb]
	\includegraphics[scale=0.32]{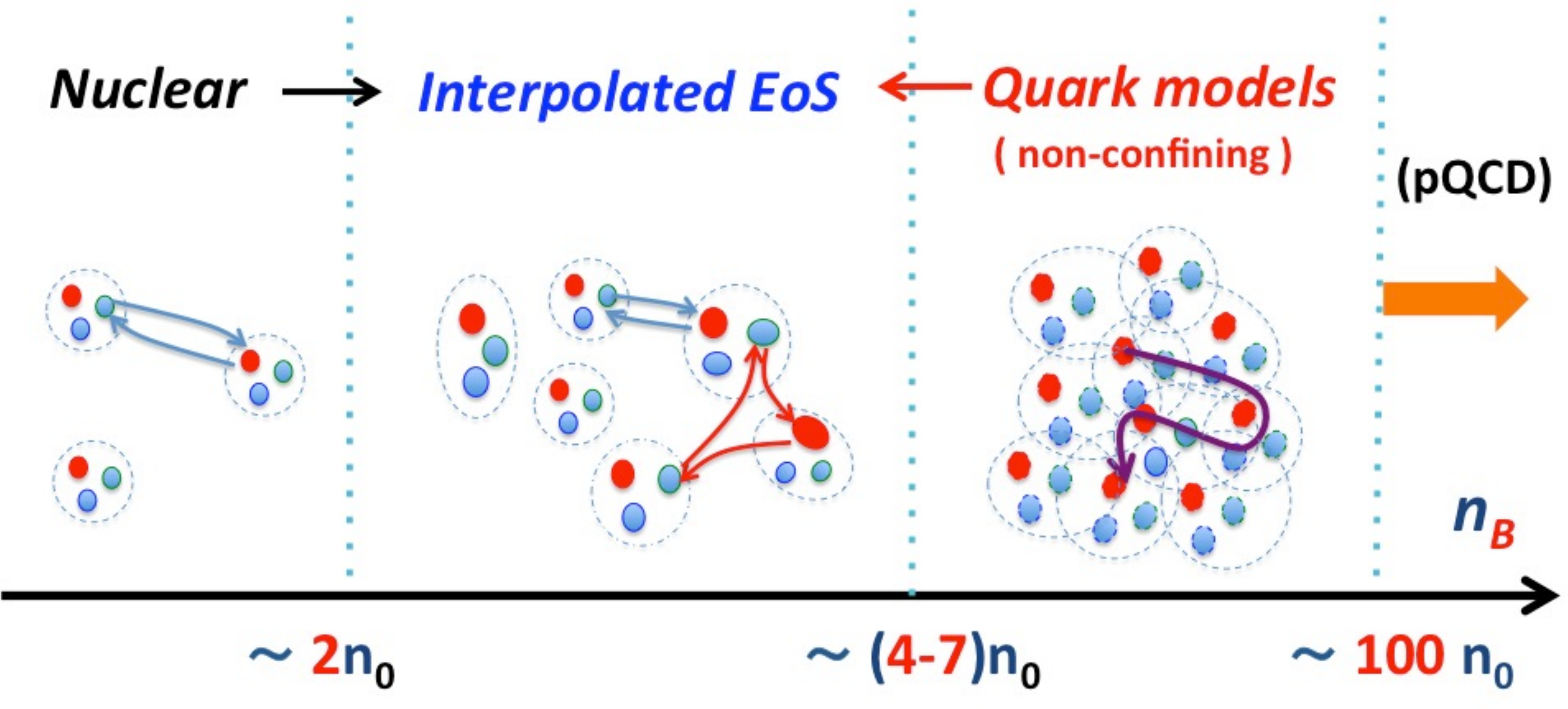}
	\caption{The idea of interpolating between known EoS of nuclear matter at low densities and deconfined quark matter at high densities is to bridge the gap in our quantitative knowledge of the physical mechanisms that govern quark deconfinement and its interplay with chiral symmetry restoration, color superconductivity and other collective phenomena in the strongly coupled quark-gluon plasma.
	Quark exchange processes between baryons are illustrated as well as the emergence of dense quark matter with itinerant, delocalised quarks.	
		From Ref.~\cite{Kojo:2014rca}.}
	\label{fig:interpol}
\end{figure}

\subsection{Interpolation}
\label{ssec:interpol}

The method of interpolating\index{interpolation method} hadron and quark matter EoS has been suggested for the neutron star EoS by Masuda, Hatsuda and Takatsuka in \cite{Masuda:2012kf,Masuda:2012ed} in analogy to an earlier developed technique for vanishing baryochemical potential along the temperature axis of the QCD phase diagram\index{QCD phase diagram} \cite{Asakawa:1995zu}.
While the first construction was flawed because it was not performed in natural thermodynamic variables, the revised work corrected for this and the method has found extensive use
\cite{Kojo:2014rca,Blaschke:2013rma} as a tool to bridge the gap in our knowledge about the nonperturbative domain of the hadron-to-quark matter transition before a reliable unified approach\index{unified EoS} based on quark degrees of freedom will be developed. 
The situation is illustrated in Fig.~\ref{fig:interpol}, taken from \cite{Kojo:2014rca}.

\begin{figure}[!htb]
	\includegraphics[scale=0.32]{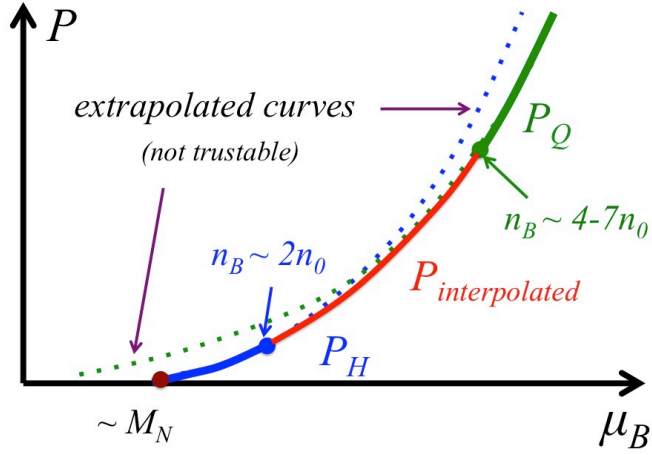}
	\caption{Illustration of the interpolation scheme\index{interpolation method} in the pressure vs. baryochemical potential diagram. This method shall be applied when a Maxwell construction makes no sense because the low-density and high-density forms of the equation of state loose their applicability before they would meet. 
    In this case the curves with the higher pressure which should be the physical ones are the not trustable parts of the given EoS. 
    From Ref.~\cite{Kojo:2014rca}.}
	\label{fig:interpol2}
\end{figure}

This method can be contrasted to the Maxwell construction and its modification which was described before.
In applying the Maxwell construction we tacitly assume that the two EoS to be matched have a common
domain of validity in the space of the thermodynamic variables so that their direct matching makes sense.
However, this may not be the case when, the low-density and high-density forms of the equation of state become not trustworthy before they would meet in the $P-\mu$ plane. 
For example when a soft nuclear matter EoS gets extrapolated to densities where already quark exchange effects should play a role and a perturbative QCD EoS is drawn into the low density region around the nuclear matter density where it is lacking confinement, chiral symmetry breaking etc. 
This is illustrated in Fig.~\ref{fig:interpol2}.
Note by comparing with the Maxwell construction case of Fig.~\ref{fig:pasta} that the interpolated part of the EoS lies now below the (unphysical) matching of the asymptotic EoS\footnote{If one would not pay attention and apply a Maxwell construction\index{Maxwell construction} to this example one would have to let the physical pressure go along the larger pressure, which would mean to join the two not trustable parts of the model EoS.
A recent example for matching unphysical EoS is Ref.~\cite{Annala:2017tqz}, where the authors in such grounds suggest a new type of holographic hybrid stars, with nuclear matter core surrounded by a quark matter shell.  
}.

For a systematic discussion of interpolation techniques in obtaining hybrid star EoS from given hadronic and quark matter model inputs, see Ref.~\cite{Alvarez-Castillo:2018pve}. 

\subsection{Metastable hadronic stars coexisting with strange quark stars}
So far we have considered hadron-to-quark matter phase transitions in compact stars in equilibrium for which the Gibbs conditions of phase equilibrium apply.
However, as it has been discussed above in subsect.~\ref{ssec:delta},
the nucleation of strange quark matter from a metastable hadronic phase by quantum tunneling \cite{Iida:1998pi,Bombaci:2006cs}
could take sufficiently long time for the metastable hadronic star\index{metastable hadronic stars} branches of compact stars to become populated and observable, together with the final state of the conversion process, the branch of stars consisting of absolutely stable strange quark matter\index{strange quark matter}. 
This is called the two-family scenario\index{two families scenario} 
\cite{Berezhiani:2002ks,Drago:2004vu,Bombaci:2004mt}
the simultaneous existence of very compact ($R<10$ km) hadronic stars with a maximum mass 
\index{maximum mass of neutron stars}
of only $1.5-16$ M$_\odot$ with less compact ($R>10.5$ km) strange quark matter stars fulfilling the 2~M$_\odot$ mass constraint could be explained.
While the conversion of a hadronic star to a strange quark star starting from a seed droplet of strange quark matter is completed within milliseconds by turbulent combustion \cite{Herzog:2011sn,Pagliara:2013tza},
the timescale for the seed formation by quantum tunneling varies between those timescales and irrelevant periods beyond the age of the Universe. 
That such a huge variation results from small changes in parameters of the model, like the bag constant\index{bag constant} of the strange quark matter EoS limits the predictive power of such a scenario.

Nevertheless, the two-family scenario has recently been applied also in interpreting the merger event GW170817\index{GW170817} that was first observed by the LIGO detectors as a chirped gravitational wave signal and subsequently as a gamma-ray burst by the FERMI and INTEGRAL satellites as well as in all other wavelengths of the electromagnetic spectrum 
\cite{TheLIGOScientific:2017qsa}.
In Refs.~\cite{Drago:2017bnf,Drago:2018nzf} 
it has been suggested that between the one-family and two-family scenarios could be discriminated once the gravitational wave signal at the moment of the merger could be detected.

\subsection{Towards a unified quark-nuclear EoS}

We have promised to come back to the issue of a unified EoS\index{unified EoS}. This time not for crust and core matter, but rather a unified description of quark and nuclear matter on the basis of a microscopic quark model where nucleons appear as bound states of quarks. 
As in the case of the unified EoS for nuclear matter and clustered nuclear matter, the point is to describe strongly coupled deconfined quark matter and the quark bound states in medium on the basis of the same model for the quark interaction.

In an early, nonrelativistic potential model approach \cite{Ropke:1986qs} this problem has been solved in these steps: (i) solve the three-quark Schr\"odinger equation for the nucleon ground state with a confining potential, (ii) determine the density-dependent nucleon self-energy (Pauli blocking\index{quark Pauli blocking}) shift from the quark exchange interaction in the two-nucleon system (antisymmetrisation of the 6-quark wave function) in the nucleonic matter phase, (iii) determine the self-consistent Hartree shift in the quark matter phase when the confining interaction is saturated within the range of the nearest neighbors (string-flip model),\index{string-flip model (SFM)} (iv) Maxwell construction\index{Maxwell construction} of the phase transition between the Fermi gas of nucleons with repulsive (quark exchange) interactions and the self-consistent Hartree approximation for the quark plasma with saturated confinement interaction.     
The result is a unified EoS for quark-nuclear matter with a first-order deconfinement phase 
transition where the description of both phases is based on the same confining interaction potential. 

The main flaw of the old string-flip model is that it works with constituent quarks and thus does not account for the chiral symmetry restoration which we expect to occur in the vicinity of the deconfinement one. 

An alternative route to the unified EoS is to start from the NJL model\index{Nambu--Jona-Lasinio (NJL) model} as a paradigmatic field theoretic model for dynamical chiral symmetry breaking and its restoration in a hot dense medium. 
This model must then be considered beyond the mean-field level in order to describe baryons as bound states 
of three quarks. 
Technically, a bosonization of the NJL model (\ref{eq:NJL_Lagrangian}) in the meson and diquark channels is performed and a Faddeev equation for the nucleon as a quark-diquark bound state can be derived from a partial resummation of a subclass of diagrams in the one-quark loop expansion, see \cite{Cahill:1988zi,Reinhardt:1989rw}. 
This scheme has been carried out for $T=0$ and finite baryochemical potential in Ref.~\cite{Bentz:2001vc} 
to arrive at an EoS for nuclear matter with composite nucleons coupled to the scalar and vector meson mean fields, thus being reminiscent of a Walecka model. 
The saturation was obtained at too high densities but could be brought to the phenomenological value by adopting asd hoc an 8-quark interaction or by imposing an infrared cutoff to the quark propagators which should mimic confinement. Subsequently, the phase transition to NJL model quark matter was obtained from a Maxwell construction \cite{Bentz:2002um} thus fulfilling the aim to obtain a unified quark-nuclear matter EoS within one microscopic model for the quark interactions. 
This NJL model based approach was extended  in Ref.~\cite{Lawley:2006ps} to the case of isospin-asymmetric hybrid star matter in $\beta-$equilibrium with a phase transition to color superconducting quark matter where also mass-radius diagrams for hybrid compact star sequences have been obtained. 
Unfortunately, Sarah Lawley left Physics and this promising development was discontinued.    

Another promising step has been done in Ref.~\cite{Wang:2010iu} where in a slightly simplified quark-diquark interaction scheme of the NJL model type the dissociation of the baryon in the QCD phase diagram with 2SC phase has been described in terms of the baryon spectral function. Interestingly, the baryon in the medium is a Borromean state: as a three-quark state it is bound while the two-quark state (diquark) is unbound, see also the discussion in \cite{Blaschke:2015sla}. Unfortunately, this development did not lead to an EoS of quark-nuclear matter yet.  
This step can be expected from a cluster virial expansion for quark matter on the basis of the $\Phi-$derivable approach\index{$\Phi-$derivable approach} which is equivalent to a generalized Beth-Uhlenbeck approach\index{Beth-Uhlenbeck approach} when the choice of diagrams in the $\Phi-$functional is restricted to all two-loop diagrams that can be drawn with quark cluster Green's functions \cite{Blaschke:2015bxa,Bastian:2018wfl}.

As drawback of the NJL model based approach remains the lack of confinement. This becomes particularly severe when the EoS is required not only at zero temperature, as in the case of supernova collapse and neutron star merger simulations. In these situations with temperatures of the order of the Fermi energy
there are free quarks contributing to the thermodynamics of the system in the hadronic phase. 
The coupling to the Polyakov-loop\index{Polyakov-loop} suppresses quarks at finite temperatures and provides an improvement but is not applicable at low and vanishing temperatures for high baryon densities. 
In this situation a relativistic density functional approach to quark matter can provide a solution which suppresses colored states in the confinement domain at low densities by diverging scalar self-energies (masses). 
An EoS for the finite-temperature applications of this concept to heavy-ion collisions has been developed in \cite{Khvorostukin:2006aw}. 
The finite-temperature generalization of the SFM EoS has been applied to supernova simulations in 
\cite{Fischer:2017lag}, where it was demonstrated that with this EoS the QCD phase transition can trigger a successful explosion for massive progenitors such as blue supergiant stars as massive as 50 M$_\odot$.
These very promising developments at the mean-field level \cite{Kaltenborn:2017hus,Khvorostukin:2006aw} have now to be followed by developing a relativistic cluster virial expansion on this basis \cite{Bastian:2018wfl}.

\section{Hybrid stars in the mass-radius diagram}
\label{sec:HSP}

It is known that  there is a one-to-one relationship between the mass-radius ($M-R$) relationship of 
(cold) compact stars and their EoS given, e.g., as pressure-energy density ($P(\varepsilon)$) relation
via the Tolman--Oppenheimer--Volkoff (TOV) equations\index{Tolman--Oppenheimer--Volkoff (TOV) equations}.
On this basis, measurements of mass and radius of pulsars can be used to determine the EoS of strongly 
interacting matter in $\beta-$equilibrium in a domain of densities and temperatures that is not accessible to terrestrial laboratory experiments.
Naturally, the question appeared whether neutron star observations could help to settle the question whether there is at least one critical point in the QCD phase diagram\index{QCD phase diagram}.
To this end one would have to prove that the deconfinement phase transition that eventually occurs in neutron star interiors is of first order.    
Exploiting the above relation between EoS and M-R diagram the task can be rephrased: 
Is there a feature in the possible M-R sequences of (hybrid) compact stars that would signal a first order phase transition in the corresponding EoS?
If this question could be answered affirmatively, what are then the prospects to identify such a feature by
observations of compact stars?
The following subsections will discuss answers to these two questions.
 
\subsection{Phase diagram of hybrid stars}

\begin{figure}[!tb]
	\includegraphics[scale=0.35]{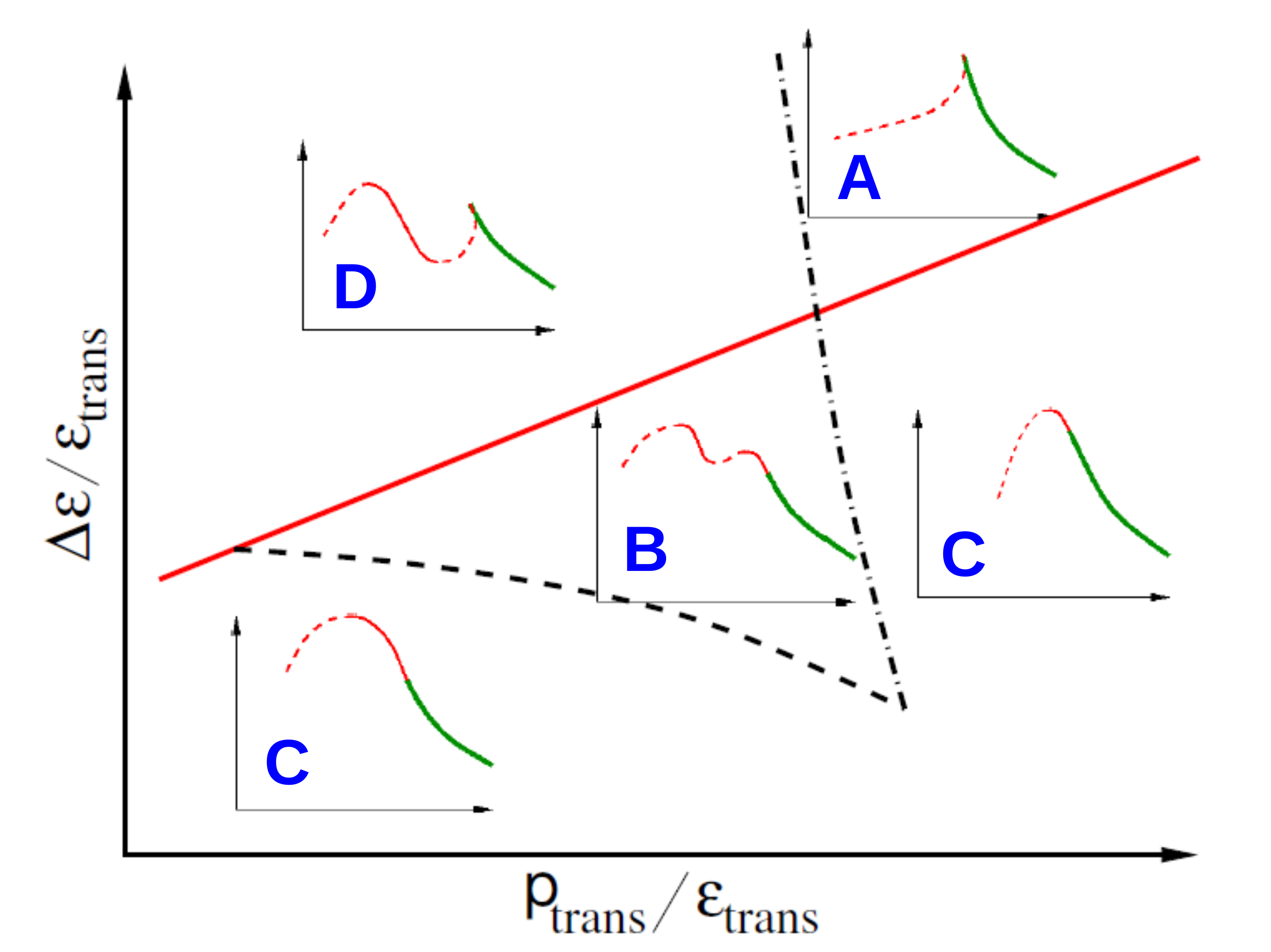}
	\caption{Classification of mass-radius sequences in the so-called phase diagram for hybrid compact stars\index{hybrid compact stars}, spanned by the latent heat $\Delta \varepsilon$ and the critical pressure $p_{\rm trans}$ of their underlying equation of state, normalized to the critical energy density $\varepsilon_{\rm trans}$ 
		at the onset of the phase transition.
		Adapted from Ref.~\cite{Alford:2013aca}. 
	}
	\label{fig:transitions}
\end{figure}

In Ref.~\cite{Alford:2013aca} Alford et al. have introduced the notion of a phase diagram for hybrid stars\index{hybrid compact stars},
i.e. stars where the core matter undergoes a phase transition of first order\index{first order phase transition} with a jump in the energy
density $\Delta \varepsilon$ (latent heat) occurring when a critical pressure $P_{\rm trans}$ is reached at a critical energy density $\varepsilon_{\rm trans}\rm trans$.
This phase diagram is spanned by latent heat and critical pressure, both in units of the energy density at
the onset of the transition, see Fig.~\ref{fig:transitions}.
It exhibits a universal dividing line (shown as solid line in red color) which corresponds to the 
criterion for a gravitational instability at the onset of the phase transition and is given by
\cite{Seidov:1971} 
\begin{eqnarray}
\label{seidov}
\frac{\Delta\varepsilon}{\varepsilon_{\rm crit}} \ge \frac{1}{2} 
+ \frac{3}{2} \frac{P_{\rm crit} }{\varepsilon_{\rm crit}} ~,
\end{eqnarray}
so that above this line stable hybrid stars are absent (A) or, if the high-density phase of matter in the inner core is stiff enough, a disconnected (D) "third family"  \cite{Gerlach:1968zz}
sequence of stable hybrid stars emerges.
The position of the (almost vertical) dividing line is not universal, it depends on the EoS.
Below the Seidov line, the onset of the phase transition in the inner core is followed by a connected (C) branch of stable hybrid stars. In this domain there is a special, triangular-shaped region where after the connected branch of hybrid stars an instability occurs which divides it from a third family sequence of compact hybrid stars, so that in this region the sequence is characterized by both (B): a connected and a disconnected branch of stable hybrid stars.       
 
The existence of a disconnected hybrid star branch in the $M-R$ diagram\index{mass-radius ($M-R$) relationship} can thus be identified as the distinctive and observationally accessible feature for detecting a (strong) first-order phase transition in compact star interiors.
Due to the mass defect which accompanies the compactification because of the gain in gravitational binding energy, there is a certain range in masses for which pairs of compact stars exist that have the same mass but different radii and different internal composition, the so-called "mass twins"\index{mass twins} \cite{Glendenning:1998ag}.
Their existence is equivalent to the existence of a third family and therefore to the strong first order phase transition in the stellar core.  
 
\subsection{The story of the twin stars}

The possibility of a third family of compact stars\index{third family of compact stars} and its relation to a phase transition in stellar matter was recognized by Gerlach \cite{Gerlach:1968zz} already in 1968, and after a while discussed by K\"ampfer 
\cite{Kampfer:1981yr} in the context of pion condensation and quark matter in compact stars. After another 17 years, the phenomenon was reconsidered by Glendenning and Kettner \cite{Glendenning:1998ag} who coined the term "twins" for the effect accompanying the third family. 
Schertler et al. \cite{Schertler:2000xq} and Bhattacharyya et al. \cite{Bhattacharyya:2004fn} took up the idea that twin stars may signal a phase transition in a compact star, but all these solutions predicted maximum masses which were well below the $2.1~M_\odot$ constraint \cite{Nice:2005fi} which was published in 2005 (and taken back in 2007), so that mass twins were put aside, also under the impression of the 
"masquerade" effect \cite{Alford:2004pf}, to be discussed below. 
But the idea of mass twins as an observable signature for a phase transition was too good to be forgotten.
In order to revive it one had to demonstrate that it could be reconciled with the meanwhile discovered very precise constraints on the maximum mass of a compact star EoS, i.e., $1.97 \pm 0.04~M_\odot$ \cite{Demorest:2010bx} for PSR J1614-2230\index{PSR J1614-2230} 
\footnote{Note that this mass for PSR J1614-2230 reported in 2010 was down-corrected in 2016 to $1.947 \pm 0.018~M_\odot$ \cite{Fonseca:2016tux} and most recently to $1.908\pm 0.016~M_\odot$ \cite{Arzoumanian:2017puf}.} and $2.01 \pm 0.04~M_\odot$ \cite{Antoniadis:2013pzd} for PSR J0438+432\index{PSR J0438+432}.
Early in 2013, it could be demonstrated that high-mass twins were theoretically possible, when a realistic nuclear EoS model was combined with a constant speed of sound (CSS) model\index{constant speed of sound (CSS) model} for the high-density phase \cite{Alvarez-Castillo:2013cxa}.    
With such a hybrid star model EoS one could describe twin stars at any mass, while fulfilling the $2~M_\odot$ mass constraint, see the recent work on the classification of twin star solutions
\cite{Christian:2017jni}. 
Even a fourth family and mass triples of compact stars are possible \cite{Alford:2017qgh}.
What are the astrophysical chances to discover twin stars, at high or low mass? We discuss two possibilities. 

\subsubsection{High-mass twin\index{mass twins} stars: J1614-2230 \& J0438+432}

\begin{figure}[htb]
	\includegraphics[width=0.9\textwidth]{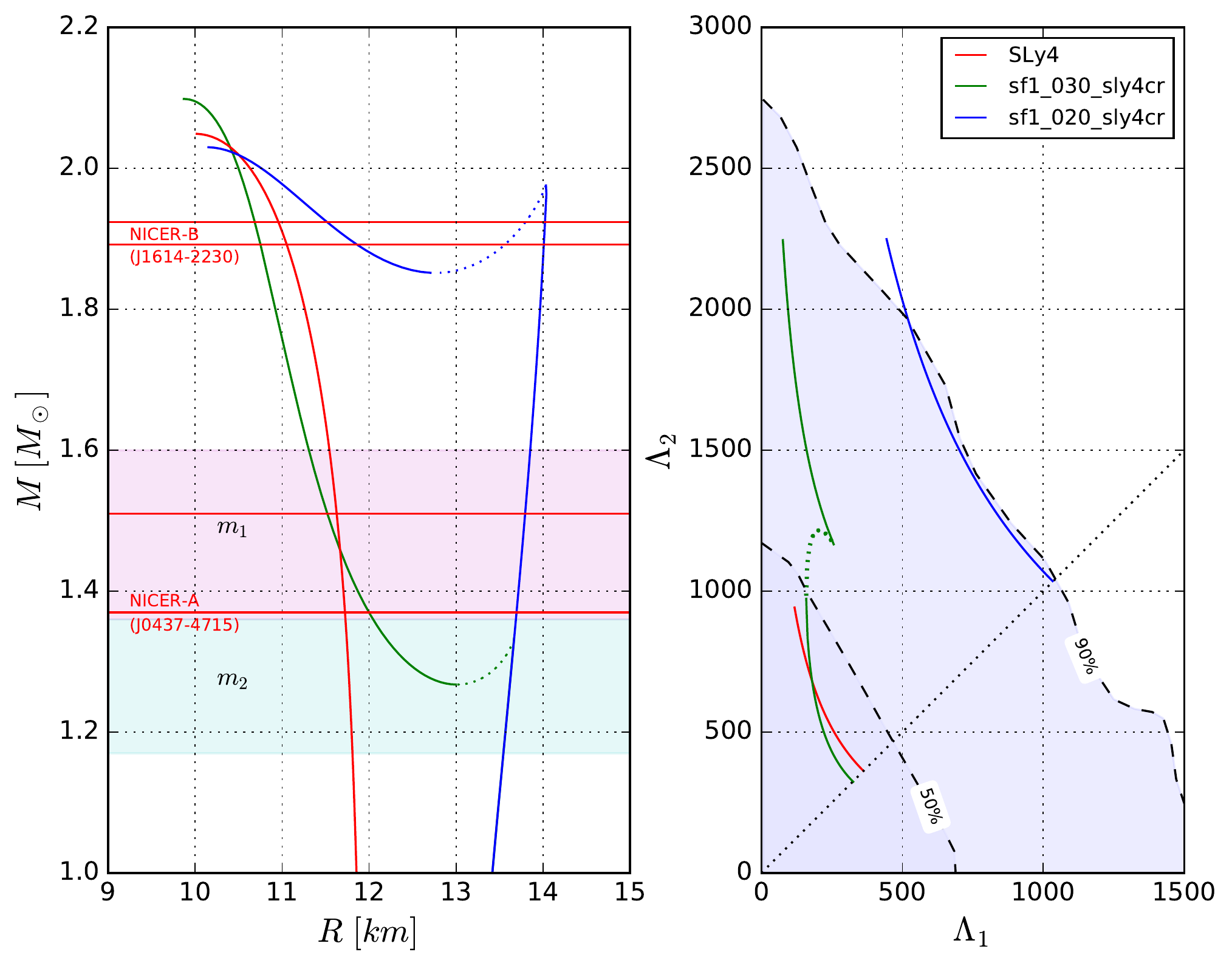}
	\caption{
    Left panel: $M-R$ relations for a stiff (DD2$\underbar{ }$p40, blue line) and a soft (Sly4, red line) hadronic EoS. The result for hybrid stars binaries with hybrid stars from the third family branch obtained by a phase transition to quark core matter (SFM with screening parameters $\alpha=0.2$,  and $0.3$, green lines) together with the mass ranges $m_1$ and $m_2$ of the stars in the binary merger GW170817.
    The red horizontal lines indicate the the mass bands for the nearest millisecond pulsar J0437-4715 \cite{Reardon:2015kba} and for the high-mass pulsar J1614-2230 \cite{Arzoumanian:2017puf}. Both pulsars are target for the radius measurement by NICER \cite{nicer}.
    Right panel:  LVC constraint \cite{TheLIGOScientific:2017qsa} on tidal deformabilities $\Lambda_2$ vs. $\Lambda_1$ for the low-spin prior with a 50\% (grey region, bordered by dashed line) and a 90\% (just ashed line) confidence level region from the binary compact star merger\index{compact star merger} GW170817, compared to results shown in the left panel. A merger of two hybrid stars from the third family branch is equivalent to that of a soft hadronic EoS.
    From Ref.~\cite{Bejger:2018}, see also \cite{Paschalidis:2017qmb}. 
	}
	\label{fig:twins}
\end{figure}

Historically, as soon as the second $2~M_\odot$ pulsar was discovered \cite{Antoniadis:2013pzd}  in 2013, it was suggested that radius measurements of these stars should have the best chance to discover the case
of high-mass twins \cite{Blaschke:2013ana} and thus to prove the existence of a critical endpoint in the QCD phase diagram \cite{Benic:2014jia,Alvarez-Castillo:2016wqj}. 
Since the deconfinement transition, if possible in compact star interiors at all, should have the best chances to occur in the most massive stars, i.e., the stars with the highest possible densities in their core\footnote{Note that the transition from a configuration at the endpoint of the neutron star branch in the $M-R$ diagram to the third family branch of hybrid stars (triggered, e.g., by mass accretion or spin-down) occurs under simultaneous conservation of baryon mass and angular momentum, as has been demonstrated in \cite{Bejger:2016emu} for the case of the high-mass twin EoS \cite{Benic:2014jia}. It could therefore occur at the free-fall timescale, accompanied by a burst-type phenomenon \cite{Alvarez-Castillo:2015dqa} 
(such as a fast radio burst \cite{Falcke:2013xpa}) unlike the case discussed earlier in \cite{Glendenning:1997fy} where an angular momentum mismatch had to be compensated by, e.g., dipole radiation an estimated timescale for the transition period of $10^5$ yrs.}. 
However, while PSR J1614-2230 is a millisecond pulsar for which the neutron star interior composition explorer (NICER)\index{neutron star interior composition explorer (NICER)} \cite{nicer} in principle could provide a radius measurement (given a sufficient duration of this experiment on board of the international space station (ISS)\index{international space station (ISS)}), such a radius measurement is not possible for PSR J0438+432. 
Then one could hope for the discovery of another high-mass pulsar, e.g., by the square kilometer array (SKA)\index{square kilometer array (SKA)} \cite{ska}.
The case is illustrated in the left panel of Fig.~\ref{fig:twins} for the hybrid star EoS of Ref.~\cite{Kaltenborn:2017hus} with the DD2$\underbar{ }$p40 hadronic EoS and the SFM with screening parameter $\alpha=0.2$, where the radius difference of the twins at $2~M_\odot$ is about 3~km. 
In that figure we show also the most recent result $1.947\pm 0.018~M_\odot$ for the mass measurement of PSR J1614-2230 \cite{Arzoumanian:2017puf}. 

Note that the case of two compact star families with an overlapping range in masses but different radii has also been considered in another context, namely for the conversion
of metastable hadronic stars to strange stars \cite{Bombaci:2016xuj,Bhattacharyya:2017mdh,Dondi:2016yjl}, 
where the branch of strange stars can reach or exceed the 2~M$_\odot$ mass constraint at larger radii than the smaller and less massive hadronic stars have.
The testable consequences of this two families scenario\index{two families scenario} in the context of gravitational wave signals from compact star merger events have been recently discussed in Refs.~\cite{Drago:2017bnf,Drago:2018nzf}.

\subsubsection{Low-mass twin stars\index{mass twins}: GW170817 \& NICER}

With the detection of gravitational waves from the inspiral phase of the neutron star merger GW170817\index{GW170817} by the LIGO Scientific and Virgo Collaboration\index{LIGO Scientific and Virgo Collaboration} 
\cite{TheLIGOScientific:2017qsa} and the subsequent detection of the associated kilonova event in the galaxy NGC4993 in all ranges of the electromagnetic spectrum \cite{Monitor:2017mdv}, the window for multi-messenger astronomy has been opened.
It is possible that already this merger event with stars from the mass ranges $M_1=1.36 - 1.60~M_\odot$ and 
$M_2=1.16 - 1.36~M_\odot$ could have been a binary not of two regular neutron stars but rather a neutron star and a hybrid star or even a pair two hybrid stars, see the right panel of Fig.~\ref{fig:twins}.
With the same hybrid EoS as for the high-mass twin case we can obtain also typical mass (low-mass) twins, just by slightly varying the unknown screening parameter to be $\alpha=0.3$.
From examining the LIGO constraint on tidal deformabilities in the right panel of Fig.~\ref{fig:twins}, it is clear that a neutron star - neutron star scenario for a soft hadronic EoS (SLy4) can not be distinguished from a hybrid star - hybrid star scenario. 
A hybrid star - neutron star scenario is also possible, but a neutron star - neutron star scenario with a stiff hadronic EoS is only marginally compatible with GW170817, see \cite{Paschalidis:2017qmb}.
In \cite{Annala:2017llu} the constraint on tidal deformability has been translated to a limiting radius for a 1.4~M$_\odot$ neutron star (within the neutron star - neutron star scenario)
of $R_{1.4, {\rm max}}=13.6$ km.

In Fig.~\ref{fig:twins} we show also the mass range $1.44 \pm 0.07~M_\odot$ of  PSR J0437-4715\index{PSR J0437-4715} \cite{Reardon:2015kba} which is the primary target of the radius measurement by NICER\index{neutron star interior composition explorer (NICER)} \cite{nicer}. While at this moment the low-mass twin case is only an option, competing with the case of an ordinary neutron star merger with soft hadronic EoS, the soft hadronic EoS could be ruled out should NICER \cite{nicer} announce a radius for the nearest millisecond pulsar PSR J0437-4715 in excess of $\sim 13~$km \cite{Paschalidis:2017qmb}.
 
\subsection{Masquerade\index{masquerade} ?}

It may well be that the deconfinement phase transition in the dense matter EoS does not leave a 
recognizable imprint on the mass-radius diagram of compact stars. 
This could have different reasons, for instance: (i) the critical density for the onset is too high to be reached in compact star interiors, even at highest masses, (ii) the quark matter EoS is too soft to carry a hadronic mantle so that the onset of quark matter results in gravitational instability and thus determines
the value of the maximum mass of neutron stars, (iii) the transition is too weak (the latent heat $\Delta \varepsilon$ is too small) to lead to a recognizable (D)isconnected third family branch but rather results in a hybrid star branch (C)onnected to the neutron star one and not distinguishable from the pure neutron star case.   

This latter case has been dubbed "masquerade" in Ref.~\cite{Alford:2004pf}.
It would require other methods of detecting the deconfinement transition in neutron star interiors than to
measure the mass-radius dependence. 
For instance the cooling behaviour could be qualitatively different for stars above the threshold mass.
The best case, however, would be a galactic supernova event which would emit a detectable second neutrino burst that has to be associated with the conversion of nuclear matter to quark matter.


\section{Conclusions}
\label{sec:conclusions}

With their tremendous gravity and their very high magnetic fields, the interior of neutron stars exhibits novel phases that cannot be reproduced in terrestrial laboratories. The atmosphere and the surface of a neutron star can be probed by spectroscopic measurements, although the inferred composition may still depend on the adopted atmospheric model. The interior of a neutron star is not directly observable. The outer most region is predicted to consist of a solid crust made of fully ionised nuclei embedded in a highly degenerate electron gas. Under the assumption of cold catalysed matter, nuclei are arranged on a perfect body-centred cubic lattice, and their composition is completely determined by experimental atomic mass measurements, currently up to a density of about $6\times 10^{10}$~g~cm$^{-3}$. Future measurements will allow us to drill deeper into the crust. However, the cold catalysed matter hypothesis may not be very realistic. The composition of a newly formed hot neutron star is likely to become frozen when the crust solidifies, or at even earlier times. Which nuclides would thus be produced and how would they arrange? Could the crust consist of a disordered solid? Following the composition of a matter element as it cools down is very challenging as this requires the knowledge of all the relevant reaction rates, in addition to the dense-matter properties that govern the thermal evolution. Moreover, the presence of a high magnetic field or the accretion of matter from a companion star may radically change the constitution and the structure of the crust.  Different neutron stars may thus have different crusts depending on their history. With increasing pressure, nuclei become progressively more neutron rich until some neutrons become unbound. The onset of neutron emission by nuclei, at densities of order $10^{11}$~g~cm$^{-3}$ marks the transition to the inner crust, where neutron-proton clusters are immersed in a neutron ocean (superfluid at low enough temperatures). With further compression, clusters might fuse and form a liquid mantle of nuclear ``pastas'' that eventually dissolve into a uniform mixture of nucleons and electrons at a density of order $10^{14}$~g~cm$^{-3}$. 

At supersaturation densities, heavier baryons such as hyperons and deltas may be excited before the deconfined quark matter phases appear, most likely in a superconducting state. 
The details of this transition are largely unknown because no benchmark exist. 
So we describe the main challenging questions that our present understanding faces and try some answers. 
What is the confinement/deconfinement mechanism in cold degenerate matter? 
How is it intertwined with chiral symmetry breaking/restoration in the presence of strong diquark correlations (color superconductivity)? 
Which mechanisms determine the stiffness of nuclear matter and quark matter? 
How could eventually a unified description of quark and nuclear matter be achieved? 
Do hyperons occur in neutron stars or does nature choose to deconfine strongly interacting matter before 
heavy baryons would get excited? Can hybrid stars with strange quark matter content be stable?
Does strangeness occur in compact star interiors at all? 
How far is the asymptotic perturbative QCD description from the density range probed by the phenomenology  
of compact stars? 
 
We discussed the existence of strange matter in compact stars in hadronic and quark matter phases.
The appearance of hyperons leads to a hyperon puzzle in approaches based on effective 
baryon-baryon potentials but is not a severe problem in relativistic mean field models.
The puzzle is resolved for a stiffening of hadronic matter at supersaturation densities, 
an effect based on the quark Pauli quenching between hadrons.
We further outlined the conflict between the necessity to implement dynamical chiral symmetry breaking 
for a realistic quark matter model and the condition of undressed, approximate massless quarks for the appearance of absolutely stable strange quark matter. 
The existence of absolutely stable strange quark matter cannot be excluded on theoretical grounds only.
However, we outlined the problems of the reasoning that lead to this hypothesis.
In general, the role of strangeness in compact stars in hadronic or quark matter realizations remains unsettled.

As a workhorse for the theory chiral quark models of the NJL type have been discussed and their extension to a relativistic density-functional theory has been advertised because it allows to implement features of quark confinement simultaneously with dynamical chiral symmetry breaking, color superconductivity and stiffening at high densities which all are required for the phenomenological description of compact star physics. Further unknown white spots in the theory of superdense neutron star matter can be circumvented with interpolating models and Bayesian techniques. 
One of the modern questions with a chance to be answered by next generations of observational campaigns is the nature of the deconfinement transition. Is it a strong first order transition which would give rise to the phenomenon of mass twin stars?

\begin{acknowledgement}
The work of N.C. was supported by Fonds de la Recherche Scientifique - FNRS (Belgium) under grants n$^\circ$~CDR-J.0187.16 and CDR-J.0115.18. 
D.B. received support from Narodowe Centrum Nauki - NCN (Poland) under contract No. 
UMO-2014/13/B/ST9/02621 (Opus7). 
This work was also partially supported by the European Cooperation in Science and Technology 
(COST) Action MP1304 \emph{NewCompStar}. 
\end{acknowledgement}

\end{document}